\documentclass[journal,draftcls,onecolumn,12pt,twoside]{IEEEtran}
\normalsize

\usepackage{lineno}

\usepackage{cite}

\ifCLASSINFOpdf
\else
\fi

\usepackage{amsmath}
\usepackage{amssymb}
\usepackage{mathrsfs}
\usepackage{multirow}

\newcounter{defcnt}
\newtheorem{lemma}{Lemma}
\newtheorem{theorem}{Theorem}

\newenvironment{pf}{\par{\it{Proof:}}\ }{\hfill $\square$\par}
\newenvironment{definition}{
	\refstepcounter{defcnt}
	\par{\it{Definition \arabic{defcnt}:}}\ }{}
\usepackage{algorithm}
\usepackage{algorithmic}
\usepackage{float}
\usepackage{lipsum}
\usepackage{xcolor}
\DeclareMathOperator{\FFT}{FFT}
\DeclareMathOperator{\IFFT}{IFFT}
\newcommand{\X}{\mathbb{X}}
\newcommand{\F}{\mathbb{F}}
\DeclareMathOperator{\rank}{rank}

\renewcommand{\Im}{\operatorname{Im}}

\makeatletter
\newenvironment{breakablealgorithm}
{
	\begin{center}
		\refstepcounter{algorithm}
		\hrule height.8pt depth0pt \kern2pt
		\renewcommand{\caption}[2][\relax]{
			{\raggedright\textbf{\ALG@name~\thealgorithm} ##2\par}%
			\ifx\relax##1\relax 
			\addcontentsline{loa}{algorithm}{\protect\numberline{\thealgorithm}##2}%
			\else 
			\addcontentsline{loa}{algorithm}{\protect\numberline{\thealgorithm}##1}%
			\fi
			\kern2pt\hrule\kern2pt
		}
	}{
		\kern2pt\hrule\relax
	\end{center}
}
\makeatother

\newtheorem{aq}{Author Query/Comment}

\newcommand{\baq}{\begin{aq}}
\newcommand{\eaq}{\end{aq}}

\begin{document}
%
\title{New Decoding of Reed--Solomon Codes Based on FFT and Modular Approach}
%
%
%

\author{Nianqi Tang and Yunghsiang S. Han, \IEEEmembership{Fellow, IEEE}
\thanks{N. Tang is with Huawei Technologies Co., Ltd., and Y. S. Han is with the Shenzhen Institute for Advanced Study, University of Electronic Science and Technology of China.}
}

\maketitle
\begin{abstract}
Decoding algorithms for Reed--Solomon (RS) codes are of great interest for both practical and theoretical reasons. 
In this paper, an efficient algorithm, called the modular approach (MA), is devised for solving the  Welch--Berlekamp (WB) key equation. By taking the MA as the key equation solver, we propose a new decoding algorithm for systematic RS codes. {For $(n,k)$ RS codes, where $n$ is the code length and $k$ is the code dimension,} the proposed decoding algorithm has both the best  asymptotic computational complexity $O(n\log(n-k) + (n-k)\log^2(n-k))$ and  the smallest constant factor achieved to  date. By comparing the number of field operations required, we show that when decoding practical RS codes, the new algorithm is significantly superior to {the} existing methods in terms of computational complexity. When decoding the $(4096, 3584)$ RS code {defined} over $\F_{2^{12}}$, the new algorithm is 10 times faster than a conventional syndrome-based method.  
Furthermore, the new algorithm has a regular architecture and is thus suitable for hardware implementation.

\end{abstract}

%
\IEEEpeerreviewmaketitle

\section{Introduction}

Reed--Solomon (RS) codes, first proposed in~\cite{Polynomial1960}, are the most commonly used error correcting codes and have been widely applied in a variety of communication systems, including storage devices, digital television, and data transmission. Research into the decoding of RS codes is therefore  of both practical and theoretical importance. {Among the  algorithms currently available for decoding RS codes, the most widely known  is  syndrome-based RS decoding}, in which the key equation is solved using either the Berlekamp--Massey (BM) algorithm or the Euclidean algorithm. For an $(n,k)$ RS code, {where $n$ is the code length and $k$ is the code dimension}, the computational complexity of syndrome-based decoding  is $O(n(n-k) + (n-k)^2)$ ({see~\cite{Error2005, Algebraic2003, lin2001error} for more details}). {Here, the computational complexity of an algorithm is expressed using the asymptotic notation $O$, where  $O(p(\varepsilon))$ denotes the set of functions
$O(p(\varepsilon)) = \{q(\varepsilon): \text{there exist positive constants}\ c\ \text{and}\ \varepsilon_0\ \text{such that}\ 0 \leq q(\varepsilon)\leq cp(\varepsilon)\ \text{for all}\ \varepsilon > \varepsilon_0\}$ in which $c$ is called the constant factor. {Note that $q(\varepsilon)\in O(p(\varepsilon))$ represents the real computational complexity of the algorithm or an upper bound on it.}
}
Another decoding algorithm with complexity $O(n(n-k) + (n-k)^2)$ was presented in~\cite{Error1984}, where the syndrome and the error locator polynomial are related by the Welch--Berlekamp (WB) key equation, and the WB algorithm is used for solving this equation. 

{Much effort has been devoted to  designing decoding algorithms with lower complexity  by using the fast Fourier transform (FFT) over finite fields. Fedorenko and Trifonov~\cite{fedorenko2002finding} proposed an algorithm for finding roots of polynomials over finite fields that exploits a specific polynomial called the $p$-polynomial. Based on this algorithm, Lin \emph{et al.}\cite{lin2007fast} then presented a fast algorithm for the syndrome calculation. Wu \emph{et al.}~\cite{wu2012reduced} used the partial composite cyclotomic Fourier transform (CFT) to derive fast syndrome-based decoders.  Bellini \emph{et al.}~\cite{bellini2011structure} proposed a method to reduce the number of additions required in the CFT, and Fedorenko~\cite{fedorenko2019efficient} further reduced the multiplicative complexity of the partial inverse CFT. 
Gao and Mateer~\cite{gao2010additive} devised an additive FFT algorithm based on a Taylor expansion. Based on the well-known subspace polynomials, Lin \emph{et al.}~\cite{FFT2016} proposed an efficient additive FFT and devised a decoding algorithm with complexity $O(n\log(n-k) + (n-k)\log^2(n-k))$ for $(n,k)$ RS codes, which is the best asymptotic computational complexity achieved to  date.}

{After the breakthrough work of Guruswami and Sudan \cite{Guruswami1999improved}, decoding RS codes beyond half of the minimum distance has drawn much attentions. 
Interpolation algorithms for solving the key equation of the Guruswami-Sudan (GS) algorithm were proposed in \cite{o2002grobner, lee2008list, Alekhnovich2005linear, beelen2010key, chowdhury2015faster}.}
{On the other hand, one-pass Chase decoding algorithms were proposed in \cite{wu2012fast} and \cite{shany2022a}, which share computations among the hard-decision decodings of the different test error patterns.}



In this paper, we devise an efficient algorithm,  the modular approach (MA), for solving the  WB key equation. We then show that this approach can be applied to {solve} the key equation proposed in~\cite{FFT2016}, and we derive a new decoding algorithm for {the} $(n,k)$ RS codes. Two versions of the MA are presented. The first,  the frequency-domain modular approach (FDMA), updates only two polynomials in the frequency domain with complexity ${O}((n-k)^2)$. It is suitable for decoding short codes. The second,  the fast modular approach (FMA), processes in a divide-and-conquer style and has a complexity ${O}((n-k)\log^2(n-k))$ for arbitrary $n-k$. We show that the new decoding algorithm has both the best  asymptotic computational complexity and  the smallest constant factor achieved to  date. We compare the proposed decoding algorithm with {the} existing methods by counting the number of field operations. The results show that the new algorithm is significantly superior to these other techniques. More precisely, for a $(4096, 3584)$ RS code, the new algorithm is 10 times faster than conventional syndrome-based RS decoding. Furthermore, this new algorithm has a regular architecture and it is therefore suitable for practical implementation.

The remainder of this paper is organized as follows. Section~\ref{sec:MA} provides a detailed discussion of the MA. A new decoding algorithm for RS codes is then proposed in Section~\ref{sec:newdecoding}. Next, we compare the new algorithm with other methods from the literature in Section~\ref{sec:comparison}. Finally, we conclude the paper in Section~\ref{sec:conclusion}.


\section{Modular Approach}\label{sec:MA}
In this section, we describe the MA, which is capable of solving the WB key equation. 
The WB key equation problem can be expressed as follows: Find polynomials $W(x)$ and $N(x)$ with $\deg(N(x)) < \deg(W(x))$ satisfying
\begin{equation}\label{eq:WB_key}
	N(x_i) = W(x_i)y_i,\quad {i = 1, 2, \dots, \rho}
\end{equation}
for a given set of nonzero points $(x_i, y_i), {i = 1, 2, \dots, \rho}$, over a field $\F_{2^m}$, where $W(x), N(x)\in \F_{2^m}[x]$. {Note that for decoding RS codes, we have $\rho = 2t$, where $t$ is the error correction capability}. Without loss of generality, we assume that the $x_i$ are distinct. 

\begin{definition}
	The rank of an ordered polynomial pair $(W(x), N(x))$ is defined as
	\begin{equation*}
		\rank[W(x), N(x)] = \max\{2\deg(W(x)), 1 + 2\deg(N(x))\}.
	\end{equation*}
\end{definition}
\noindent Note that the rank of a polynomial pair containing  a zero polynomial  is dominated by its nonzero component, and we then define  $\rank[0,0] = 0$. 
It has been shown that there exist two complementary solutions $(W(x), N(x))$ and $(V(x), M(x))$ of problem (\ref{eq:WB_key}) such that
\begin{gather*}
	\rank[W(x), N(x)] + \rank[V(x), M(x)] = 2\rho + 1 \mbox{ and }
	W(x)M(x) - V(x)N(x) = c\prod_{i = 1}^{\rho}(x - x_i)
\end{gather*}
for some nonzero scalar $c\in\F_{2^m}$. We definitely  have $\rank[W(x), N(x)] \neq \rank[V(x), M(x)]$, since $2\rho + 1$ is odd. Among the two complementary solutions, the one with lower rank is desired for decoding RS codes. 
It should be mentioned that the  definition of rank presented here uses a different polynomial order from the original definition in~\cite{Error1984} (more precisely, it uses $(W(x), N(x))$ instead of $(N(x), W(x))$), since it is convenient to have the same order as the basis matrix defined below. More detailed discussions can be found in~\cite{Error2005, Error1984, morii1992generalized}. 

By characterizing the solution set of the WB key equation as {an} $\F_{2^m}[x]$-module, the so-called modular approach provides an efficient algorithm for constructing the desired solution. Before presenting the MA, we  first review some essential concepts regarding modules and homomorphisms. 

\begin{definition}\label{def:module}
	For the polynomial ring $\F_{2^m}[x]$,  an $\F_{2^m}[x]$-module $\mathcal{Q}$ is an abelian group with a law of composition, written as $+$, together with a scalar multiplication $\F_{2^m}[x] \times\mathcal{Q}\to\mathcal{Q} $, written as $a,v\rightarrow av$, that satisfies the  axioms
	\begin{equation}\label{eq:module}
	1v = v, \quad (ab)v = a(bv), \quad (a+b)v = av + bv, \quad a(v + v' ) = av + av' 
	\end{equation}
	for all $a,b\in \F_{2^m}[x]$, $v, v' \in\mathcal{Q}$ and such that the  results of the operations  in~\eqref{eq:module} are still in $\mathcal{Q}$.
\end{definition}

Notice that these are precisely the axioms for a linear space except that the scalars come from a ring rather than a field. Thus, modules are  natural generalizations of linear spaces to rings. Hence, the concepts of basis and independence can be carried over from linear spaces to modules. However, the number of elements of a basis for a module is called the rank, instead of the dimension.

\begin{definition}\label{def:rank_mod}
	An $\F_{2^m}[x]$-module $\mathcal{Q}$ is called a free $\F_{2^m}[x]$-module of rank $2$ if there exist independent elements $v, v' \in \mathcal{Q}$ such that any $w\in \mathcal{Q}$ can be represented as a linear combination of $v$ and $v' $, i.e., $w = av + bv' $ for $a,b\in \F_{2^m}[x]$. The set $\{v, v' \}$ is called a basis of $\mathcal{Q}$.
\end{definition}


Next, the concept of a module homomorphism is introduced so that the solution set to the WB key equation can be described as the kernel of a specific module homomorphism.

\begin{definition}
	For  two $\F_{2^m}[x]$-modules $\mathcal{Q}$ and  $\mathcal{Q}' $, a module homomorphism $\varphi: \mathcal{Q}\to \mathcal{Q}' $ is a map that is compatible with the laws of composition:
	$\varphi(v + v' )  = \varphi(v) +  \varphi(v' )\quad  \text{and}\quad \varphi(av) = a \varphi(v)$
	for all $v, v' \in \mathcal{Q}$ and $a\in \F_{2^m}[x]$. The kernel of $\varphi$, denoted by $\ker(\varphi)$, is the set of elements of $\mathcal{Q}$ that are mapped to the additive identity $0$ in $\mathcal{Q}' $, i.e., $\ker(\varphi) = \{v\in \mathcal{Q}\mid \varphi(v) = 0 \}$.
\end{definition}

Detailed discussions of modules and homomorphisms can be found in a variety of modern algebra books; see, for example,~\cite{herstein2006topics}. 

We rewrite the WB key equation~\eqref{eq:WB_key}  in a more general form as
\begin{equation}\label{eq:WB_general}
	d_iW(x) + g_iN(x) \equiv 0 \pmod{x - x_i},\quad i = 1, 2, \dots, \rho,
\end{equation}
by setting $d_i = y_i$ and  $g_i = 1$. For each $i$, we define a module homomorphism {$\phi_i:\F_{2^m}[x]^2\to\F_{2^m}$} on the corresponding equation in~\eqref{eq:WB_general} by
\begin{align*}
	\phi_i(W(x), N(x)) &= d_iW(x) + g_iN(x) \bmod{(x - x_i)}
	=(W(x_i), N(x_i))\begin{pmatrix}
	d_i\\
	g_i
	\end{pmatrix}.
\end{align*}
Note that  $\F_{2^m}[x]^2$ represents the module of $\F_{2^m}[x]$-vectors, i.e., row vectors with entries in $\F_{2^m}[x]$. Clearly, $\F_{2^m}[x]^2$ is a free $\F_{2^m}[x]$-module of rank $2$ with a basis $\{(1,0), (0,1)\}$. The kernel of $\phi_i$ characterizes the solution to the $i$th equation of~\eqref{eq:WB_general}, and the intersection 
$\ker(\phi_1)\cap\ker(\phi_2)\cap\dots\cap\ker(\phi_{\rho})$
is the solution to the set of congruences {in}~\eqref{eq:WB_general}. 
Next, as $x - x_i, i = 1, 2, \dots, \rho$, are pairwise relatively prime, if we define the homomorphism $\phi:\F_{2^m}[x]^2\to\F_{2^m}[x]$ by
\begin{equation}\label{eq:gen_hom}
	\phi(W(x), N(x)) =D(x)W(x) + G(x)N(x) \bmod{\Pi(x)},
\end{equation}
where $D(x_i) = d_i, G(x_i) = g_i$ and $\Pi(x) = \prod_{i = 1}^{\rho}(x - x_i)$, then it follows that 
$\ker(\phi) = \ker(\phi_1)\cap\ker(\phi_2)\cap\dots\cap\ker(\phi_{\rho})$
by the Chinese remainder theorem. Note that we can make the assumption that the greatest common divisor $\gcd(D(x), G(x))$ is relatively prime to $\Pi(x)$. 

From the above discussion, the solution set to the WB key equation~\eqref{eq:WB_key} is exactly the kernel of $\phi$. Next, we demonstrate that $\ker(\phi)$ is a free  $\F_{2^m}[x]$-module of rank $2$ and that an irreducible basis matrix, which is desired for decoding RS codes, exists in $\ker(\phi)$.
Before describing $\ker(\phi)$, we first develop the concept of a basis matrix. 
Hereinafter, {\bf the notation $\mathcal{Q}$ represents a free $\F_{2^m}[x]$-module of rank $2$ satisfying $\mathcal{Q}\subseteq\F_{2^m}[x]^2$}.

\begin{definition}
	$\Psi$ is called a basis matrix of $\mathcal{Q}$ if its rows form a basis of $\mathcal{Q}$. 
\end{definition}

\begin{theorem}[\cite{Fast1995}]\label{the:basis}
	$\ker(\phi)$ is a free $\F_{2^m}[x]$-module of rank $2$ and the followings hold:
	\begin{enumerate}
		\item For any basis matrix $\Psi$ of $\ker(\phi)$, we have
		$\det(\Psi) = c\Pi(x)$ for some nonzero $c\in \F_{2^m}$.
		\item Conversely, if the rows of $\Phi \in \F_{2^m}[x]^{2\times2}$ are in $\ker(\phi)$ and $\det(\Phi) = c\Pi(x)$ for some nonzero $c\in \F_{2^m}$, then $\Phi$ is a basis matrix of $\ker(\phi)$.
	\end{enumerate}
	
\end{theorem}

%

Although $\ker(\phi)$ can be described by an arbitrary basis matrix, the one with the lowest complexity is desired for decoding RS codes. The complexity of a basis matrix is characterized by a property called irreducibility. 

\begin{definition}
	{
	A basis matrix 
	$\begin{pmatrix}
	W(x) & N(x)\\
	V(x) & M(x)
	\end{pmatrix}$
	of $\ker(\phi)$ is said to be irreducible if, for any basis matrix $\begin{pmatrix}
	W^{'}(x) & N^{'}(x)\\
	V^{'}(x) & M^{'}(x)
	\end{pmatrix}$, we have
	\begin{equation*}
		\rank[W(x), N(x)] + \rank[V(x), M(x)] \leq \rank[W^{'}(x), N^{'}(x)] + \rank[V^{'}(x), M^{'}(x)].
	\end{equation*}}
\end{definition}

\begin{lemma}[\cite{Error2005}]\label{lem:irre}
	{
	A basis matrix 
	$\begin{pmatrix}
	W(x) & N(x)\\
	V(x) & M(x)
	\end{pmatrix}$
	of $\ker(\phi)$ is irreducible if
	$\rank[W(x), N(x)] + \rank[V(x), M(x)] = 2\rho + 1.$	}
	\end{lemma}


{According to Lemma \ref{lem:irre}, if we find a basis matrix which satisfies the rank constraint in Lemma \ref{lem:irre}, then its rows are solutions to the WB key equation with the lowest complexity. Hence, the WB key equation problem is converted to constructing a basis matrix of $\ker(\phi)$ which satisfies the rank constraint.}
We now turn to describe the MA, which is an efficient algorithm for finding such a matrix. 
The main idea behind the MA, roughly speaking, is to find a module chain step by step. For example, if we want to solve $\ker(\phi_1)\cap\ker(\phi_2)$, the MA first constructs an irreducible basis matrix of $\ker(\phi_1)$. Next, the homomorphism $\phi_2$ is updated and restricted to $\ker(\phi_1)$ so that a desired solution can be found. As  $\ker(\phi_1)\cap\ker(\phi_2)$ is a free $\F_{2^m}[x]$-module of rank $2$, by Theorem~\ref{the:basis},  $\ker(\phi_1)\cap\ker(\phi_2)\subseteq\ker(\phi_1)$, which forms a module chain. 
Before describing the MA precisely, we introduce the concept of a projection of a module.

\begin{definition}
	{A homomorphism $\psi:\F_{2^m}[x]^2\to\mathcal{Q}$ is called a projection of $\mathcal{Q}$ if and only if the image of $\F_{2^m}[x]^2$ under $\psi$ is equal to $\mathcal{Q}$, i.e., $\Im (\psi) = \mathcal{Q}$.}
\end{definition}

	
	{\textit{Remarks:} Let $\Psi$ be any basis matrix of $\mathcal{Q}$. Then the homomorphism $\psi:\F_{2^m}[x]^2\to\mathcal{Q}$ defined by $\psi(W(x), N(x)) = (W(x), N(x))\Psi$
	is a projection of $\mathcal{Q}$.}

The following lemma provides a powerful tool for finding the module chain. 

\begin{lemma}\label{lem:update}
	Let $\psi$ be a projection of $\mathcal{Q}$ and let $\varphi$ be a homomorphism that maps $ \F_{2^m}[x]^2$ to $\F_{2^m}[x]^2$. Then 
	$\mathcal{Q}\cap\ker(\varphi) = \psi(\ker(\varphi\circ\psi)),$
	where {$\varphi\circ\psi$ denotes  composition of maps, with $\psi$ being applied first, followed by $\varphi$}.
	
	\begin{pf}
		Note that $\psi(\ker(\varphi\circ\psi))\subseteq \Im (\psi) = \mathcal{Q}$. Since $\varphi$ is a module homomorphism, by definition, $\varphi(\psi(\ker(\varphi\circ\psi))) = 0$ such that $\psi(\ker(\varphi\circ\psi))\subseteq \ker(\varphi)$. Then 
		$\psi(\ker(\varphi\circ\psi))\subseteq \mathcal{Q}\cap\ker(\varphi)$.
		
		Conversely, $\varphi\circ\psi(\psi^{-1}(\mathcal{Q}\cap\ker(\varphi)))) = {0}$,  which implies $\psi^{-1}(\mathcal{Q}\cap\ker(\varphi)) \subseteq  \ker(\varphi\circ\psi)$. It follows that
		$\mathcal{Q}\cap\ker(\varphi) \subseteq \psi(\ker(\varphi\circ\psi))$.
		Hence, one must have
		$\mathcal{Q}\cap\ker(\varphi) = \psi(\ker(\varphi\circ\psi))$.
	\end{pf}
\end{lemma}

We are now ready to describe the MA. 
For $i\in \{1, 2, \dots, \rho\}$ and $j\in \{0, 1, \dots, \rho\}$, recursively define the homomorphisms ${\phi}_{i}^{j}:\F_{2^m}[x]^2\to\F_{2^m}$  by
\begin{equation*}
	{\phi}_{i}^{j}(W(x), N(x)) = \phi_{i}((W(x), N(x))\Psi_1^j),
\end{equation*}
where
\begin{equation*}
\Psi_1^j =
\begin{cases}
\text{identity matrix}\ I_{2\times2}, &j = 0,\\
\Psi_j\Psi_{j-1}\cdots\Psi_{1}, & j > 0,
\end{cases}
\end{equation*}
and $\Psi_j$ is a basis matrix of $\ker(\phi_j^{j-1})$ for $j > 0$. For any matrix $\Psi\in\F_{2^m}[x]^{2\times2}$,  $\Psi(x_i)$ denotes the matrix whose terms are the evaluations of the corresponding terms of $\Psi$ at $x_i$. 
\begin{lemma}
	${\phi}_{i}^{j}$, $\Psi_{1}^{j}$, and $\Psi_j$ are well-defined.
	\begin{pf}
		Because $\Psi_1^0$ is an identity matrix, we have $\phi_1^0 = \phi_1$, which implies that $\Psi_1$ exists according to Theorem~\ref{the:basis}. Now suppose that $\Psi_1, \Psi_2, \dots, \Psi_{j - 1}$ exist. Then $\Psi_1^{j - 1}$ must exist, which means that
		\begin{align*}
			\phi_{j}^{j - 1}(W(x), N(x)) &= \phi_{j}((W(x), N(x))\Psi_1^{j-1})
			= (W(x_j), N(x_j))\Psi_1^{j-1}(x_j)
			\begin{pmatrix}
			d_j\\
			g_j
			\end{pmatrix}
		\end{align*}
		Since $\phi_j^{j-1}$ is a special case of  $\phi$, by a similar proof to that of Theorem~\ref{the:basis}, we have that $\ker(\phi_j^{j-1})$ is a free $\F_{2^m}[x]$-module of rank $2$. Hence, $\Psi_j$ exists by Theorem~\ref{the:basis}. 
		
		Since $\Psi_j$ exists for $j = 1, 2, \dots, \rho$,  $\Psi_1^j$ is well-defined, which also implies that $\phi_i^j$ is well-defined for all $i = 1, 2, \dots, \rho$ and $j = 0, 1, \dots, \rho$.
	\end{pf}
\end{lemma}

For $i\in \{1, 2, \dots, \rho\}$ and $j\in \{0, 1, \dots, \rho\}$, we rewrite the homomorphism $\phi_i^j$ as
\begin{equation*}
{\phi}_i^j(W(x), N(x)) = (W(x_i), N(x_i))\Psi_{1}^{j}(x_i)
\begin{pmatrix}
{d}_i\\
{g}_i
\end{pmatrix}
= (W(x_i), N(x_i))
\begin{pmatrix}
{d}_i^j\\
{g}_i^j
\end{pmatrix}
\end{equation*}
and  define the homomorphisms $\psi_j, \psi_1^j:\F_{2^m}[x]^2\to\F_{2^m}[x]^2$ by
\begin{align*}
	\psi_j(W(x), N(x)) = (W(x), N(x))\Psi_j \mbox{ and } 
	\psi_1^j(W(x), N(x)) = (W(x), N(x))\Psi_1^j.
\end{align*}
It is easy to see that $\psi_1^j = \psi_1^{j-1} \circ \psi_j$ and  $\phi_i^j = \phi_i \circ \psi_1^j$.
\begin{lemma}\label{lem:bar_phi}
	${\phi}_j^{j-1}$ is nontrivial\footnote{A trivial homomorphism maps all elements to the additive identity.} for $j = 1, 2, \dots, \rho$.
	\begin{pf}
		The proof is by induction on $j$. As $\phi_1^0 = \phi_1$,  the claim is true for $j = 1$ by the assumption that $\phi_1$ is nontrivial. Next, suppose that for $l = 1, 2, \dots, j - 1$, ${\phi}_l^{l-1}$ are nontrivial. It follows that $\det(\Psi_l) = c_l(x - x_l)$ and that $\det(\Psi_1^{j - 1}) = \prod_{l = 1}^{j-1}c_l(x-x_l)$ for nonzero scalars $c_l$. If ${\phi}_j^{j-1}$ is trivial, one must have $\det(\Psi_1^{j - 1}(x_j)) = 0$, which is impossible since $x_1, \dots, x_j$ are distinct. Hence, we can conclude that ${\phi}_j^{j-1}$ is nontrivial for $j = 1, 2, \dots, \rho$.
	\end{pf}
\end{lemma}
As $\phi_j^{j-1}$ is nontrivial, either $d_j^{j-1}$ or $g_{j}^{j-1}$ is nonzero.
Next, we show that, by suitably choosing $\Psi_j$, $\Psi_1^j$ is an irreducible basis matrix of $\ker(\phi_1)\cap\cdots\cap\ker(\phi_{j})$.
Define the map $R:\F_{2^m}[x]^{2\times2}\to \{0,1\}$ as
	\begin{equation*}
		R(\Psi_1^j) =
		\begin{cases}
		1& \text{if the
			first row of $\Psi_1^j$ has a larger rank than its second row},\\
		0& \text{otherwise}.
		\end{cases}
	\end{equation*}

\begin{lemma}
	\label{lem:basis_mat}
	Let the basis matrix $\Psi_j$ be equal to
		 \begin{align}\label{term 1)}
		&\begin{pmatrix}
		-g_j^{j-1} & {d}_j^{j-1} \\
		x - x_j & 0
		\end{pmatrix}
		&&\text{if}\  g_j^{j-1} = 0\ \text{or}\ ({d}_j^{j-1}\neq 0\ \text{and}\ R(\Psi_1^{j-1}) = 0)\\
\intertext{and to}
	         \label{term 2)}
		&\begin{pmatrix}
		-g_j^{j-1} & {d}_j^{j-1} \\
		0 & x - x_j
		\end{pmatrix}
		&&\text{otherwise}.
		\end{align}
	Then 
	$\Psi_{1}^j =
	\begin{pmatrix}
	W(x) & N(x)\\
	V(x) & M(x)
	\end{pmatrix}$
	satisfies $\rank[W(x), N(x)] + \rank[V(x), M(x)] = 2j + 1$, and $\Psi_{1}^j$ is an irreducible basis matrix of $\ker(\phi_1)\cap\ker(\phi_2)\cap\cdots\cap\ker(\phi_j)$.
	
	\begin{pf}
		The proof is by induction on $j$.
		For $j = 1$, we have ${\phi}_1^0 = \phi_1$. Since $\Psi_1^0 = I$,  $R(\Psi_1^{0}) = 0$.	
		When ${g}_1^0 = 0$ or $(d_1^0\neq 0\ \text{and}\ R(\Psi_1^{0}) = 0)$, $\Psi_1$ is equal to the matrix in~\eqref{term 1)}.
		It is straightforward to see that $(-{g}_1^0, {d}_1^0)$ and $(x - x_1, 0)$ are included in $\ker(\phi_1)$ by applying ${\phi}_1$ to them. Thus, $\Psi_1$ is a basis matrix of $\ker(\phi_1)$ by Theorem~\ref{the:basis}, since $\det(\Psi_1) = d_1^0(x - x_1)$. Next, we have
		$\rank[-{g}_1^0, d_1^0] + \rank[x - x_1, 0] = 1 + 2 = 3$.
		Therefore, by Lemma~\ref{lem:irre}, $\Psi_1^1 = \Psi_1$ is an irreducible basis matrix of $\ker({\phi}_1)$.
		
		On the other hand, it is straightforward to verify the claims for the case in which $\Psi_1$ is equal to the matrix in~\eqref{term 2)} by repeating the above proof.
		
		
		We have proved the claims for $j = 1$. Next, suppose that the claims are true for $1, 2, \dots, j - 1$. 
		
		For all cases, it is easy to verify that $\Psi_j$ is a basis matrix of $\ker({\phi}_j^{j-1})$   by repeating the above proof.
		By induction, $\Psi_1^{j - 1}$ is an irreducible basis matrix of $\ker(\phi_1)\cap\cdots\cap\ker(\phi_{j-1})$. So, $\psi_1^{j-1}$ is a projection of $\ker(\phi_1)\cap\cdots\cap\ker(\phi_{j-1})$. According to Lemma~\ref{lem:update}, we have
		\begin{align*}
		\ker(\phi_1)\cap\cdots\cap\ker(\phi_{j-1})\cap\ker(\phi_{j}) &=  \psi_1^{j-1}(\ker(\phi_j \circ \psi_1^{j-1}))
		= \psi_1^{j-1}(\ker({\phi}_j^{j-1}))\
		= \psi_1^{j}(\F_{2^m}[x]^2),
		\end{align*}
		where the last equality follows because $\psi_j$ is a projection of $\ker({\phi}_j^{j-1})$.
		Then the polynomial pairs $\psi_1^j(1,0)$ and $\psi_1^j(0,1)$, which are exactly the two rows of $\Psi_1^{j}$, are contained in $\ker(\phi_1)\cap\cdots\cap\ker(\phi_{j})$. Furthermore, as $\det(\Psi_{1}^j) = \det(\Psi_{j})\det(\Psi_1^{j - 1}) = c\prod_{l = 1}^{j}(x - x_l)$
		for some nonzero $c\in\F_{2^m}$,  $\Psi_1^{j}$ is a basis matrix of $\ker(\phi_1)\cap\cdots\cap\ker(\phi_{j})$ by Theorem~\ref{the:basis}.
		
		It remains to show that $\Psi_1^{j}$ is irreducible.
		We write $\Psi_1^{j - 1}$ as
		$\begin{pmatrix}
		W' (x) & N' (x)\\
		V' (x) & M' (x)
		\end{pmatrix}.$ 
		Recall that 
		$\rank[W' (x), N' (x)] + \rank[V' (x), M' (x)] = 2(j - 1)$.

		When ${g}_j^{j-1} = 0$ or $(d_j^{j-1}\neq 0\ \text{and}\ R(\Psi_1^{j-1}) = 0)$, 
		\begin{align*}
		\Psi_1^j =
		\begin{pmatrix}
		W(x) & N(x)\\
		V(x) & M(x)
		\end{pmatrix}
		= \begin{pmatrix}
		-{g}_j^{j-1} & d_j^{j-1} \\
		x - x_j & 0
		\end{pmatrix}
		\begin{pmatrix}
		W' (x) & N' (x)\\
		V' (x) & M' (x)
		\end{pmatrix}.
		\end{align*}
		If ${g}_j^{j-1} = 0$ or $R(\Psi_1^{j-1}) = 0$, then it is easy to verify that 
		\begin{align*}
		\rank[W(x), N(x)] &= \rank[-{g}_j^{j-1}W' (x) + d_j^{j-1}V' (x), -{g}_j^{j-1}N' (x) + d_j^{j-1}M' (x)]\\
		&= \rank[V' (x), M' (x)].
		\end{align*}
		As $\rank[V(x), M(x)] = \rank[(x-x_j)W' (x), (x-x_j)N' (x)] = \rank[W' (x), N' (x)] + 2$, it follows that $
		\rank[W(x), N(x)] + \rank[V(x), M(x)] = 2(j - 1) + 2 = 2j + 1$. Then
		$\Psi_1^{j}$ is an irreducible basis matrix of $\ker(\phi_1)\cap\cdots\cap\ker(\phi_{j})$ by Lemma~\ref{lem:irre}.
		
		On the other hand, if
		\begin{align*}
		\Psi_1^j =
		\begin{pmatrix}
		W(x) & N(x)\\
		V(x) & M(x)
		\end{pmatrix}
		= \begin{pmatrix}
		-{g}_j^{j-1} & d_j^{j-1} \\
		0 & x - x_j
		\end{pmatrix}
		\begin{pmatrix}
		W' (x) & N' (x)\\
		V' (x) & M' (x)
		\end{pmatrix},
		\end{align*}
		then, by   preceding as before, one can verify that
		$\rank[W(x), N(x)] + \rank[V(x), M(x)]= 2j + 1$. By Lemma~\ref{lem:irre}, $\Psi_1^{j}$ is irreducible in $\ker(\phi_1)\cap\cdots\cap\ker(\phi_{j})$. This completes the proof.
	\end{pf}
\end{lemma}

A detailed description of the MA is presented in Algorithm~\ref{alg:MA}, in which $r_1^j$ and $r_2^j$ denote the ranks of the two rows of $\Psi_1^j$. {Since $\Psi_1^0 = I_{2\times2}$, we  have $r_1^0 = 0, r_2^0 = 1$ at the beginning. At the $j$th iteration, the MA first identifies the basis matrix $\Psi_j$ of $\ker({\phi}_j^{j-1})$ by checking the conditions ${g}_j^{j-1} \neq 0 $, ${d}_j^{j-1} \neq 0 $, and $r_1^{j-1} < r_2^{j-1}$. Then $\Psi_j(x_i)$ can be obtained by simply substituting $x_i$ into $\Psi_j$. Next, ${d}_i^{j}$, ${g}_i^{j}$, and the basis matrix $\Psi_1^j$ are updated in parallel. Finally, the MA returns an irreducible basis matrix $\Psi_1^t$ for decoding RS codes.}

\begin{breakablealgorithm}
	\caption{Modular Approach}
	\label{alg:MA}
	\begin{algorithmic}[1]
		\REQUIRE
		{$\{x_i, d_i, g_i\}, i = 1, 2, \dots, \rho$.}
		\ENSURE
		{An irreducible basis matrix $\Psi_1^{\rho}$ of $\ker(\phi_1)\cap\ker(\phi_2)\cap\cdots\cap\ker(\phi_{\rho})$, $r_1^{\rho}, r_2^{\rho}$,
		where $\phi_i, i = 1, 2, \dots, \rho$ are homomorphisms defined by $\phi_i(W(x), N(x)) = d_iW(x_i) + g_iN(x_i)$.}
		
		\STATE {\textbf{Initialization:}
		${d}_i^0 = d_i, {g}_i^0 = g_i$ for $i = 1, 2, \dots, \rho$, $\Psi_1^0 = I_{2\times 2}$, $r_1^0 = 0, r_2^0 = 1$.}
		
		\FOR {$j = 1, 2, \dots, \rho$}
		\IF {${g}_j^{j-1} = 0$ or ($d_j^{j-1}\neq0$ and $r_1^{j-1} < r_2^{j-1}$)}
		\STATE
		Let
		\begin{equation*}
			\Psi_j =
			\begin{pmatrix}
			-{g}_j^{j-1} & {d}_j^{j-1}\\
			x - x_{j} & 0
			\end{pmatrix}
			\ \text{and}\ 
			\Psi_j(x_i) = 
			\begin{pmatrix}
			-{g}_j^{j-1} & {d}_j^{j-1}\\
			x_{i} - x_{j} & 0
			\end{pmatrix}
			\ \text{for}\ i = 1, 2, \dots, \rho
		\end{equation*}
		\STATE $r_1^j = r_2^{j-1}, r_2^j = r_1^{j-1} + 2$.
		\ELSE
		\STATE
		Let
		\begin{equation*}
		\Psi_j =
		\begin{pmatrix}
		-{g}_j^{j-1} & {d}_j^{j-1}\\
		0 & x - x_{j}
		\end{pmatrix}
		\ \text{and}\ 
		\Psi_j(x_i) = 
		\begin{pmatrix}
		-{g}_j^{j-1} & {d}_j^{j-1}\\
		0 & x_{i} - x_{j}
		\end{pmatrix}
		\ \text{for}\ i = 1, 2, \dots, \rho
		\end{equation*}	
		\STATE $r_1^j = r_1^{j-1}, r_2^j = r_2^{j-1} + 2$.
		\ENDIF
		\FOR {$i = j + 1, j + 2, \dots, \rho$}
		\STATE
		\begin{equation*}
		\begin{pmatrix}
		{d}_i^j\\
		{g}_i^j
		\end{pmatrix}
		=
		\Psi_j(x_i)
		\begin{pmatrix}
		{d}_i^{j-1}\\
		{g}_i^{j-1}
		\end{pmatrix}
		\end{equation*}
		\ENDFOR
		\STATE
		$\Psi_1^j = \Psi_j\Psi_1^{j-1}$.
		\ENDFOR
		\RETURN $\Psi_1^{\rho}, r_1^{\rho}, r_2^{\rho}$.
	\end{algorithmic}
\end{breakablealgorithm}

The original MA was first proposed in~\cite{Fast1995}. However, there exist many differences between the original approach and the new one presented here. First, we define an irreducible basis matrix here and prove its existence for the desired kernel. Second, a more efficient algorithm is proposed for finding such an irreducible basis matrix. {The original method in~\cite{Fast1995} needs to find a homomorphism with nonzero ${d}_j^{j-1}$ during each iteration, which significantly limits its speed, especially in hardware implementation.} However, our new method  here eliminates the need for such a procedure, thereby enabling the development of a high-speed architecture, which is a prerequisite for real applications. 
Finally, the new method tracks the ranks during each iteration, which is essential for identifying uncorrectable errors.

It is well known that both the WB algorithm and the Euclidean algorithm are capable of solving the WB key equation~\eqref{eq:key_WB}. However, the WB algorithm executes the polynomial evaluations and the polynomial updates in sequence during each iteration. This means that the operations in each iteration of the WB algorithm cannot be done in parallel. Therefore, its hardware implementation has a long critical path. More details about the WB algorithm can be found in~\cite{Error1984}. On the other hand, the Euclidean algorithm fails to provide an efficient method for decoding RS codes based on the FFT, and for that reason we do not discuss it here. 

{There are several algorithms which find specific elements of a module. Fitzpatrick \cite{fitzpatrick1995on} presented a method for finding a low-weight element of the solution to the key equation $z(x)\equiv \lambda(x)\bmod{x^{2t}}$. Algorithms for solving the rational interpolation problem in the GS algorithm were proposed in \cite{o2002grobner, lee2008list, Alekhnovich2005linear, beelen2010key, chowdhury2015faster}. Compared with these algorithms, an advantage of the proposed method is that the operations in each iteration can be done in parallel, which is an important feature in implementation. Furthermore, the coming section proves that the fast modular approach is superior in terms of complexity.}
\section{Decoding Reed--Solomon Codes Based on FFT}\label{sec:newdecoding}
 
In this section, a new algorithm is presented for decoding RS codes based on the FFT, which takes the MA as the key equation solver. Two versions of the MA are presented. The first,  the frequency-domain modular approach (FDMA), is suitable for decoding short RS codes. The second,  the fast modular approach (FMA), is suitable for decoding medium or long RS codes. We shall see that the new decoding algorithm has the smallest constant factor achieved to  date, while also reaching the best known asymptotic computational complexity. 

\subsection{FFT Algorithm}
Let $(v_0,v_1,\dots,v_{m-1})$ be a basis of $\F_{2^m}$ over $\F_2$. The elements in $\F_{2^m}$ can be represented by 
\begin{align*}
\omega_l = l_0v_0 + l_1 v_1 + \dots + l_{m-1}v_{m-1}, \quad 0\leq l < 2^m,  
\end{align*}
where $l_0,\dots,l_{m-1}\in\{0,1\}$ is the binary representation of $l$. 
{The subspace polynomial is defined as
$s_{\tau}(x) = \prod_{l = 0}^{2^{\tau} - 1} (x - \omega_l)$ for $\tau = 0, 1, \dots, m$. Obviously, we have $\deg(s_{\tau}(x)) = 2^{\tau}$.}
Then the polynomial given by
\begin{equation*}
\bar{X}_l(x) = \frac{s_0(x)^{l_0}s_1(x)^{l_1}\cdots s_{m-1}(x)^{l_{m-1}}}{s_0(v_0)^{l_0}s_1(v_1)^{l_1}\cdots s_{m-1}(v_{m-1})^{l_{m-1}}}
\end{equation*}
has  degree  $l$ for $l = 0, 1, \dots, 2^m -1$. {Therefore, the set
$\bar{\mathbb{X}} = \{ \bar{X}_0(x), \bar{X}_1(x), \dots, \bar{X}_{2^m - 1}(x) \}$
is a basis of the linear space $\F_{2^m}[x]/(x^{2^m}-x)$}, which implies that any polynomial $f(x)$ in this space can be represented as a linear combination:
$f(x) = \sum_{l = 0}^{2^{m}-1}\bar{f}_l\bar{X}_l(x)$.
The vector $\bar{\mathbf{f}}=(\bar{f}_0,\bar{f}_1,\dots,\bar{f}_{2^{m}-1})$ is the coordinate vector of $f(x)$ with respect to the basis $\bar{\X}$. 

Given that $\deg(f(x)) < 2^{\tau}$, the fast Fourier transform (FFT), denoted by $\FFT_{\bar{\X}}$, evaluates $f(x)$ at points $\{\omega_l + \beta\mid l = 0, 1, \dots, 2^{\tau} - 1\}$:
\begin{align*}
\FFT_{\bar{\X}}(\bar{\mathbf{f}}, \tau, \beta)
= \mathbf{F} = (f(\omega_0 + \beta), f(\omega_1 + \beta), \dots, f(\omega_{2^\tau-1} + \beta)),
\end{align*}
for any $\beta \in \F_{2^m}$ and $\tau = 0, 1, \dots, m$, which involves ${\tau2^\tau}/{2}$ field multiplications and ${\tau2^\tau}$ field additions~\cite{Lin2014}.
The inverse FFT, denoted by $\IFFT_{\bar{\X}}$, calculates $\mathbf{\bar{f}}$ given $\mathbf{F}$, which also involves ${\tau2^\tau}/{2}$ field multiplications and ${\tau2^\tau}$ field additions in a direct implementation. Algorithms~\ref{FFTX} and \ref{IFFTX} present the details of $\FFT_{\bar{\mathbb{X}}}$ and $\IFFT_{\bar{\mathbb{X}}}$, respectively.

\begin{algorithm}[h]
	\caption{$\FFT_{{\bar{\mathbb{X}}}}$~\cite{Lin2014}}
	\label{FFTX}
	\begin{algorithmic}[1]
		\REQUIRE
		$\bar{\mathbf{f}} = (\bar{f}_0,\bar{f}_1,\dots,\bar{f}_{2^\tau-1})$, $\tau$, $\beta$.
		\ENSURE
		$(f(\omega_0 + \beta), f(\omega_1 + \beta), \dots, f(\omega_{2^\tau-1} + \beta))$.
		
		\IF {$\tau = 0$} \RETURN $\bar{f}_0$
		\ENDIF
		\FOR {$l = 0, 1,\dots,2^{\tau-1}-1$}
		\STATE {$a_l^{(0)} = \bar{f}_l + \dfrac{s_{\tau-1}(\beta)}{s_{\tau-1}(v_{\tau-1})}\bar{f}_{l+2^{\tau-1}}$}
		\STATE {$a_l^{(1)} = a_l^{(0)} + \bar{f}_{l+2^{\tau-1}}$}
		\ENDFOR
		\STATE $\mathbf{a}^{(0)}=(a_0^{(0)},\dots,a_{2^{\tau-1}-1}^{(0)}),\mathbf{a}^{(1)}=(a_0^{(1)},\dots,a_{2^{\tau-1}-1}^{(1)})$ 
		\STATE Calculate $\mathbf{A}_0 = \FFT_{\bar{\mathbb{X}}}(\mathbf{a}^{(0)},\tau-1,\beta)$, $\mathbf{A}_1 = \FFT_{\bar{\mathbb{X}}}(\mathbf{a}^{(1)},\tau-1,v_{\tau-1} + \beta)$
		\RETURN $(\mathbf{A}_0,\mathbf{A}_1)$
	\end{algorithmic}
\end{algorithm}

\begin{algorithm}[h]
	\caption{Inverse Transform of the Basis $\bar{\mathbb{X}}$~\cite{Lin2014}}
	\label{IFFTX}
	\begin{algorithmic}[1]
		\REQUIRE
		$\mathbf{F} = (f(\omega_0 + \beta), f(\omega_1 + \beta), \dots, f(\omega_{2^\tau-1} + \beta)), \tau, \beta$
		\ENSURE
		$\bar{\mathbf{f}}$ such that $ \mathbf{F}= \FFT_{\bar{\mathbb{X}}}(\bar{\mathbf{f}},\tau,\beta)$
		
		\IF {$\tau = 0$} 
		\RETURN $f(\omega_0 + \beta)$
		\ENDIF
		
		\STATE  $\mathbf{A}_{0}=(f(\omega_0 + \beta),\dots,f(\omega_{2^{\tau-1}-1} + \beta)),\mathbf{A}_{1}=(f(\omega_{2^{\tau-1}} + \beta)),\dots,f(\omega_{2^{\tau}-1} + \beta))$ 
		\STATE $\mathbf{a}^{(0)} = \IFFT_{\bar{\mathbb{X}}}(\mathbf{A}_0,\tau-1,\beta)$
		, $\mathbf{a}^{(1)} = \IFFT_{\bar{\mathbb{X}}}(\mathbf{A}_1,\tau-1,v_{\tau-1} + \beta)$
		
		\FOR {$l = 0, 1,\dots,2^{\tau-1}-1$}
		\STATE {$\bar{f}_{l+2^{\tau-1}} = a_l^{(0)} + a_l^{(1)}$}
		\STATE {$\bar{f}_l = a_l^{(0)} + \dfrac{s_{\tau-1}(\beta)}{s_{\tau-1}(v_{\tau-1})}\bar{f}_{l+2^{\tau-1}}$}
		\ENDFOR
		\RETURN $\mathbf{\bar{f}}$
	\end{algorithmic}
\end{algorithm}

{For $\mu \in\{ 0, 1, \dots, \tau\}$, if we let 
$\mathbf{F} = (\mathbf{F}_1, \mathbf{F}_2, \dots, \mathbf{F}_{2^{\tau-\mu}})$ and $\bar{\mathbf{f}} = (\bar{\mathbf{f}}_1, \bar{\mathbf{f}}_2, \dots, \bar{\mathbf{f}}_{2^{\tau-\mu}})$, where
$\mathbf{F}_i=(	f(\omega_{(i-1)2^\mu}+\beta), f(\omega_{(i-1)2^\mu+1}+\beta),\ldots,f(\omega_{i2^\mu-1}+\beta))$, $\bar{\mathbf{f}}_i=(\bar f_{(i-1)2^\mu},\bar f_{(i-1)2^\mu+1},\ldots, \bar f_{i2^\mu-1})$, then 
\cite[Lemma 10]{FFT2016} and \cite[Lemma 1]{Fast2020}
\begin{equation*}
	\IFFT_{\bar{\X}}(\mathbf{F}_1,\mu,\beta) + \IFFT_{\bar{\X}}(\mathbf{F}_2,\mu,\omega_{2^{\mu}} + \beta) + \cdots + \IFFT_{\bar{\X}}(\mathbf{F}_{2^{\tau-\mu}},\mu,\omega_{2^{\tau} - 2^{\mu}} + \beta) = \bar{\mathbf{f}}_{2^{\tau-\mu}}.
\end{equation*}}
As we shall see later, this important property is crucial for encoding and decoding RS codes. 
 
\subsection{Encoding Reed--Solomon Codes}
For an $(n, k)$ RS code where $n = 2^m$, $k = 2^m - 2^\mu $ and $\mu\in\{0, 1, \dots, m - 1\}$, the codewords are given by
{$\FFT_{\bar{\X}}(\bar{\mathbf{f}}, m, 0) = \mathbf{F} = (f(\omega_0), f(\omega_1), \dots, f(\omega_{2^m-1}))$}
for all polynomials $f(x)$ of degree  less than $2^m - 2^\mu$.
It follows that
\begin{equation*}
	\IFFT_{\bar{\X}}(\mathbf{F}_1,\mu,0) + \IFFT_{\bar{\X}}(\mathbf{F}_2,\mu,\omega_{2^{\mu}}) + \cdots + \IFFT_{\bar{\X}}(\mathbf{F}_{2^{m-\mu}},\mu,\omega_{2^{m} - 2^{\mu}}) = \bar{\mathbf{f}}_{2^{m-\mu}} = \mathbf{0}.
\end{equation*}
{If we let $\mathbf{F}_1$ be the check locations and $\mathbf{F}_2, \mathbf{F}_3, \dots, \mathbf{F}_{2^{m-\mu}}$ be the message locations}, then the encoding process is
$\FFT_{\bar{\X}}(\IFFT_{\bar{\X}}(\mathbf{F}_2,\mu,\omega_{2^{\mu}}) + \cdots + \IFFT_{\bar{\X}}(\mathbf{F}_{2^{m-\mu}},\mu,\omega_{2^{m} - 2^{\mu}}), \mu, 0)$.
The computational complexity of the encoding algorithm is ${O}(n\log(n-k))$. 
\subsection{Decoding Reed--Solomon Codes}
The received vector can be represented by
\begin{align*}
	\mathbf{r} = \mathbf{F} + \mathbf{e} = (f(\omega_0), f(\omega_1), \dots, f(\omega_{2^m-1})) + (e_0, e_1, \dots, e_{2^m - 1}),
\end{align*}
where $\mathbf{e}$ is the error pattern. If we write
$E = \{\omega_{l}\mid e_l\neq 0\mbox{ for }0\le l\le 2^m-1\}$, 
then the error locator polynomial can be defined as
$\lambda(x) = \prod_{a\in E}(x-a)$.
{Note that there exists a polynomial $r(x) \in \F_{2^m}[x]$ with degree less than $2^m$ satisfying} $
{r}(\omega_{l}) = f(\omega_l) + e_l$ for $l = 0, 1, \dots, 2^m - 1$, 
which implies that $
f(\omega_l)\lambda(\omega_l) =
r(\omega_l)\lambda(\omega_l)$.
Thus, the congruence $f(x)\lambda(x) \equiv r(x)\lambda(x) \pmod{s_m(x)}$ holds. Therefore,  there exists $q(x) \in \F_{2^m}[x]$ such that
\begin{equation}\label{eq:key_eq0}
f(x)\lambda(x) = r(x)\lambda(x) + q(x)s_m(x).
\end{equation}
{Clearly, we have $\deg(f(x)) < 2^m - 2^\mu$, $\deg(\lambda(x)) \leq 2^{\mu - 1}$, $\deg(r(x)) < 2^m$ and $\deg(s_m(x)) = 2^m$. Then the equation \eqref{eq:key_eq0} implies that $\deg(q(x)) < \deg(\lambda(x))$.
Dividing $s_m(x)$, $f(x)\lambda(x)$, and $r(x)$ by $p_{2^m-2^{\mu}}\bar{X}_{2^m - 2^{\mu}}(x)$, where $$p_{2^m - 2^{\mu}} = s_0(v_0)^{l_0}s_1(v_1)^{l_1}\cdots s_{m-1}(v_{m-1})^{l_{m-1}}$$
and $(l_0, l_1, \dots, l_{m-1})$ is the binary representation of $2^m - 2^{\mu}$, it follows that
\begin{align*}
	s_m(x) &= p_{2^m-2^{\mu}}\bar{X}_{2^m - 2^{\mu}}(x)(s_{\mu}(x) +  s_{\mu}(v_{\mu}))+ \eta_s(x)
 	,\\
 	f(x)\lambda(x) &= p_{2^m-2^{\mu}}\bar{X}_{2^m - 2^{\mu}}(x)z^{'}(x) + \eta_f(x),
	r(x) = p_{2^m-2^{\mu}}\bar{X}_{2^m - 2^{\mu}}(x)u(x) + \eta_{r}(x),
\end{align*}
where $\deg(\eta_s(x)), \deg(\eta_f(x)),\deg(\eta_r(x))$ are less than $\deg(p_{2^m-2^{\mu}}\bar{X}_{2^m - 2^{\mu}}(x))$.
When we divide both sides of~\eqref{eq:key_eq0} by $p_{2^m-2^{\mu}}\bar{X}_{2^m - 2^{\mu}}(x)$ and keep the quotients, it becomes
\begin{equation*}
z^{'}(x) = u(x)\lambda(x) + q(x)s_{\mu}(x) + s_{\mu}(v_{\mu})q(x).
\end{equation*}
As $\deg(f(x)\lambda(x)) < 2^m - 2^{\mu} +\deg(\lambda(x))$, one can conclude that $\deg(z^{'}(x)) < \deg(\lambda(x))$. Let $z(x) = z^{'}(x) - s_{\mu}(v_{\mu})q(x)$. We can then obtain the key equation:
\begin{equation}\label{eq:key_eq1}
	z(x) = u(x)\lambda(x) + q(x)s_{\mu}(x),
\end{equation}
where $\deg(z(x)) < \deg(\lambda(x))$.} Note that if the received vector $\mathbf{r}$ is a codeword, the degree of ${r}(x)$ is less than $2^m-2^{\mu}$, which implies that $u(x) = 0$. Hence, $u(x)$ can be treated as the syndrome polynomial. {Given $\mathbf{r}$, the coordinate vector of $u(x)$ with respect to $\bar{\X}$ can be computed by
\begin{equation}\label{eq:synd}
	\sum_{i = 0}^{2^{m-\mu} - 1}\IFFT_{\bar{\X}}(\mathbf{r}_i,\mu, \omega_{i\cdot 2^\mu})/p_{2^m-2^{\mu}},
\end{equation}
where $\mathbf{r}_i = (r_{i\cdot2^{\mu}}, r_{i\cdot2^{\mu} + 1}, \dots, r_{i\cdot2^{\mu} + 2^{\mu} - 1})$ is the sub-vector of $\mathbf{r}$. Details are given in \cite{FFT2016, Fast2020}.}

The key equation~\eqref{eq:key_eq1} can be rewritten as
\begin{equation}\label{eq:key_WB}
z(x) = u(x)\lambda(x)\bmod{\prod_{i = 0}^{2^{\mu} - 1}(x - \omega_i)},
\end{equation}
where $\deg(z(x)) < \deg(\lambda(x))$. {This is in the WB form and hence can be solved by the MA}.

Once the error locator polynomial $\lambda(x)$ has been obtained, its roots can be calculated by the FFT algorithm:
\begin{equation}\label{eq:FFT_search}
\FFT_{\bar{\X}}(\bar{\mathbf{\lambda}},\mu,\omega_{l\cdot{2^\mu}}), \quad l = 0, 1, \dots ,2^{m - \mu} - 1,
\end{equation}
where $\bar{\mathbf{\lambda}}$ is the coordinate vector of $\lambda(x)$ with respect to $\bar{\X}$.

It remains to compute the error values. The formal derivative of~\eqref{eq:key_eq0} is
\begin{equation*}
f' (x)\lambda(x) + f(x)\lambda' (x)
= {r}' (x)\lambda(x)+ {r}(x)\lambda' (x) + q' (x)s_m(x) + q(x).
\end{equation*}
For an error locator $\omega_l\in E$, we have
$f(\omega_l)\lambda' (\omega_l) = {r}(\omega_l)\lambda' (\omega_l) + q(\omega_l)$. 
It follows that $f(\omega_l) - {r}(\omega_l) = {q(\omega_l)}/{\lambda' (\omega_l)}$.
If $\omega_{l}$ is a message location, then, by~\eqref{eq:key_eq1}, we have
$q(\omega_l) = {z(\omega_l)}/{s_{\mu}(\omega_l)}$.
{Hence,  Forney's formula for solving the error value is}
\begin{equation}\label{eq:Forney}
f(\omega_l) - {r}(\omega_l) = \frac{z(\omega_l)}{s_{\mu}(\omega_l)\lambda' (\omega_l)}.
\end{equation}
{Note that there is no need to correct the errors in check locations.} 


Detecting uncorrectable errors is crucial in real applications. In the above decoding algorithm, a correctable error occurs if and only if
\begin{align}
& \deg(\lambda(x)) \leq {2^{\mu - 1}},\label{eq:condition-1}\\
& {\deg(z(x)) < \deg(\lambda(x)),\label{eq:condition-2}}\\
& |\{\omega_l\mid \lambda(\omega_l) = 0, l = 0, 1,\dots, 2^m-1\}| =\deg( \lambda(x)). \label{eq:condition-3}
\end{align}
Note that the MA always ensures that $\deg(\lambda(x)) \leq {2^{\mu - 1}}$, since $\rank[\lambda(x), z(x)] \leq 2^{\mu}$. In addition, if $\deg(z(x)) \geq \deg(\lambda(x))$, then $\rank[\lambda(x), z(x)]$ must be odd. Hence, tracking the ranks is enough to check the condition~\eqref{eq:condition-2}. Finally, the condition~\eqref{eq:condition-3} can be checked by the FFT algorithm~\eqref{eq:FFT_search}. As a result, all of the uncorrectable errors can be detected.

The computational complexities of computing the syndrome, finding roots of $\lambda(x)$, and  Forney's formula are ${O}(n\log(n-k))$. More detailed discussions can be found in~\cite{FFT2016,Fast2020}.

\subsection{Frequency-Domain Modular Approach}
{We now turn to solving the key equation~\eqref{eq:key_WB} using the MA. Clearly, if we set $x_i = \omega_{i - 1}$, $d_i = u(\omega_{i-1})$,  and $g_i = 1$ for $i = 1, 2, \dots, 2^\mu$, then Algorithm~\ref{alg:MA} provides two polynomial pairs satisfying the key equation, and the one with lower rank is exactly the  desired solution. However,  the FFT algorithm given in~\eqref{eq:FFT_search} requires that the polynomials to be evaluated be represented with respect to $\bar{\X}$, and therefore basis transformations are needed if the polynomials obtained are represented with respect to the monomial basis. To  avoid the need for these basis transformations, we devise the FDMA. 
The FDMA updates $\Psi_1^j(\omega_i)$ instead of $\Psi_1^j$. Note that as $\deg(z(x)) \leq 2^{\mu-1}$ and $\deg(\lambda(x)) \leq 2^{\mu-1}$,  $2^{\mu-1} + 1$ points in the frequency domain are enough for determining $\lambda(x)$ or $z(x)$, which implies that we need to update only $\Psi_1^j(\omega_i), i = 0, 1, \dots, 2^{\mu-1}$. Furthermore, because $(\lambda(x), z(x))$ satisfies the key equation~\eqref{eq:key_WB}, we must have $z(\omega_i) = u(\omega_i)\lambda(\omega_i)$ for $i = 0, 1, \dots, 2^{\mu-1}$.  Thus, the evaluations of $z(x)$ can be performed immediately once $\lambda(\omega_i)$ are available.
To sum up, the FDMA computes only the first column of $\Psi_1^{j}(\omega_i)$ during the iterations and then identifies $\lambda(\omega_i)$ by the rank. Next, it computes $z(\omega_i)$ once the iterations have been done. Finally, extended $\IFFT_{\bar{\X}}$ algorithms are used to obtain the coordinate vectors of $\lambda(x), z(x)$ with respect to $\bar{\X}$.
Algorithm~\ref{alg:FDMA} shows the details of the FDMA. {Note that we set $x_i = \omega_{i - 1}$ here.} Compared with Algorithm~\ref{alg:MA}, the FDMA computes only two polynomials in the frequency domain, rather than four. This further reduces the computational complexity and makes the FDMA suitable for hardware implementation.} 
\begin{breakablealgorithm}
	\caption{Frequency-Domain Modular Approach (FDMA)}
	\label{alg:FDMA}
	\begin{algorithmic}[1]
		\REQUIRE
		{$\{\omega_{i-1}, u(\omega_{i-1})\}, i = 1, 2, \dots, 2^{\mu}$.}
		 
		\ENSURE
		$(\lambda(x), z(x))$ that are represented with respect to $\bar{\X}$ and $\rank[\lambda(x), z(x)]$.
		
		\STATE {\textbf{Initialization:}
		${d}_i^0 = u(\omega_{i-1}), {g}_i^0 = 1$ for $i = 1, 2, \dots, 2^{\mu}$.}
		
		$W(\omega_i) = 1, V(\omega_i) = 0$ for $i = 0, 1, \dots, 2^{\mu-1}$, $r_1^0 = 0, r_2^0 = 1$.
		
		\FOR {$j = 1, 2, \dots, 2^{\mu}$}
		\IF {${g}_j^{j-1} = 0$ or (${d}_j^{j-1} \neq 0$ and $r_1^{j-1} < r_2^{j-1}$)}
		\STATE
		Let
		\begin{equation*}
		\Psi_j(\omega_i) = 
		\begin{pmatrix}
		-{g}_j^{j-1} & {d}_j^{j-1}\\
		\omega_{i} - \omega_{j-1} & 0
		\end{pmatrix}
		\ \text{for}\ i \in \{0, 1, \dots, 2^{\mu}-1\}
		\end{equation*}
		\STATE $r_1^j = r_2^{j-1}, r_2^j = r_1^{j-1} + 2$.
		\ELSE
		\STATE
		Let
		\begin{equation*}
		\Psi_j(\omega_i) =
		\begin{pmatrix}
		-{g}_j^{j-1} & {d}_j^{j-1}\\
		0 & \omega_{i} - \omega_{j-1}
		\end{pmatrix}
		\ \text{for}\ i \in \{0, 1, \dots, 2^{\mu} - 1\}
		\end{equation*}	
		\STATE $r_1^j = r_1^{j-1}, r_2^j = r_2^{j-1} + 2$.
		\ENDIF
		\FOR {$i = j + 1, j + 2, \dots, 2^{\mu}$}
		\STATE
		\begin{equation*}
		\begin{pmatrix}
		{d}_i^j\\
		{g}_i^j
		\end{pmatrix}
		=
		\Psi_j(\omega_{i-1})
		\begin{pmatrix}
		{d}_i^{j-1}\\
		{g}_i^{j-1}
		\end{pmatrix}
		\end{equation*}
		\ENDFOR
		\FOR {$i = 0, 1, \dots, 2^{\mu - 1}$}
		\STATE
		\begin{equation*}
		\begin{pmatrix}
		W(\omega_i)\\
		V(\omega_i)
		\end{pmatrix}
		=
		\Psi_j(\omega_i)
		\begin{pmatrix}
		W(\omega_i)\\
		V(\omega_i)
		\end{pmatrix}
		\end{equation*}
		\ENDFOR
		\ENDFOR
		\IF{$r_1^{2^{\mu}}>r_2^{2^{\mu}}$}
			\STATE $\lambda(\omega_i) = V(\omega_i), i = 0, 1, \dots, 2^{\mu-1}$.
		\ELSE
		\STATE $\lambda(\omega_i) = W(\omega_i), i = 0, 1, \dots, 2^{\mu-1}$.
		\ENDIF
		\STATE $z(\omega_i) = \lambda(\omega_i)d_{i + 1}^0,  i = 0, 1, \dots, 2^{\mu-1}$.
		\STATE Call Algorithm~\ref{alg:ex_IFFT} to obtain $\lambda(x)$ and $z(x)$.
		\RETURN $(\lambda(x), z(x))$ and $\rank[\lambda(x), z(x)] = \min(r_1^{2^{\mu}}, r_2^{2^{\mu}})$.
	\end{algorithmic}
\end{breakablealgorithm}

\begin{algorithm}
	\caption{Extended $\IFFT_{\bar{\X}}$}
	\label{alg:ex_IFFT}
	\begin{algorithmic}[1]
		\REQUIRE
		$f(\omega_{i}+\beta), i = 0, 1, \dots, 2^\mu$; $\mu$, $\beta$.
		\ENSURE
		$f(x)$ represented in $\bar{\X}$.
		\STATE
		Call Algorithm~\ref{IFFTX} with input $(f(\omega_0 + \beta), f(\omega_1+ \beta),\dots, f(\omega_{2^{\mu}-1}+ \beta)), \mu, \beta$ to obtain $\hat{f}(x)$
		\STATE
		Evaluate $\hat{f}(x)$ at $\omega_{2^{\mu}} + \beta$ to obtain $\hat{f}(\omega_{2^{\mu}}+\beta)$
		\STATE
		Let 
		\begin{align*}
			f(x) = &(f(\omega_{2^{\mu}}+\beta) - \hat{f}(\omega_{2^{\mu}}+\beta))\!\left(\bar{X}_{2^{\mu}}(x) - \frac{s_{\mu}(\beta)}{s_{\mu}(v_{\mu})}\right) + \hat{f}(x)\\
			= &(f(\omega_{2^{\mu}}+\beta) - \hat{f}(\omega_{2^{\mu}}+\beta))\bar{X}_{2^{\mu}}(x) + \hat{f}(x)\\
			 &- \frac{s_{\mu}(\beta)}{s_{\mu}(v_{\mu})} (f(\omega_{2^{\mu}}+\beta) - \hat{f}(\omega_{2^{\mu}}+\beta))\bar{X}_{0}(x)
		\end{align*}
		
		\RETURN
		$f(x)$
	\end{algorithmic}
\end{algorithm}

\begin{lemma}
	Given $f(\omega_{i}+\beta), i = 0, 1, \dots, 2^{\mu}, \mu$, and any $\beta\in \F_{2^m}$, Algorithm~\ref{alg:ex_IFFT} outputs the corresponding $f(x)$, and its complexity is ${O}(\mu2^{\mu})$.
	\begin{pf}
		Since $\hat{f}(x)$ is obtained by calling Algorithm~\ref{IFFTX}, it follows that $\hat{f}(\omega_i + \beta) = f(\omega_i + \beta)$ for $i = 0, 1, \dots, 2^{\mu} - 1$ and that $\deg(\hat{f}(x)) < 2^{\mu}$. Because $\bar{X}_{2^{\mu}}(x) = s_{\mu}(x)/s_{\mu}(v_{\mu})$,  we have $\bar{X}_{2^{\mu}}(\omega_i + \beta) = s_{\mu}(\omega_i)/s_{\mu}(v_{\mu}) + s_{\mu}(\beta)/s_{\mu}(v_{\mu})$. Recall that $s_{\mu}(x) = \prod_{l = 0}^{2^{\mu} - 1}(x - \omega_l)$. So, for $i = 0, 1, \dots, 2^{\mu} - 1$, we have $\bar{X}_{2^{\mu}}(\omega_i + \beta) = s_{\mu}(\beta)/s_{\mu}(v_{\mu})$. Hence, for $i = 0, 1, \dots, 2^{\mu} - 1$,
		\begin{equation}
			(f(\omega_{2^{\mu}}+\beta) - \hat{f}(\omega_{2^{\mu}}+\beta))\!\left(\bar{X}_{2^{\mu}}(\omega_i + \beta) - \frac{s_{\mu}(\beta)}{s_{\mu}(v_{\mu})}\right) + \hat{f}(\omega_i + \beta) = f(\omega_i + \beta).
		\end{equation}
		Furthermore, if $i = \omega_{2^{\mu}}$, we have
		\begin{align*}
			&(f(\omega_{2^{\mu}}+\beta) - \hat{f}(\omega_{2^{\mu}}+\beta))\!\left(\bar{X}_{2^{\mu}}(\omega_{2^{\mu}} + \beta) - \frac{s_{\mu}(\beta)}{s_{\mu}(v_{\mu})}\right) + \hat{f}(\omega_{2^{\mu}} + \beta)\\
			&= (f(\omega_{2^{\mu}}+\beta) - \hat{f}(\omega_{2^{\mu}}+\beta))s_{\mu}(v_{\mu})/s_{\mu}(v_{\mu}) + \hat{f}(\omega_{2^{\mu}} + \beta)\\
			&= f(\omega_{2^{\mu}} + \beta).
		\end{align*}
		Recall that $\bar{X}_0(x) = 1$. Therefore, Algorithm~\ref{alg:ex_IFFT} outputs the desired polynomial with respect to $\bar{\X}$. Clearly, the computational complexity of Algorithm~\ref{IFFTX} is ${O}(\mu2^{\mu})$, and the evaluation of $\hat{f}(x)$ at a single point needs ${O}(2^{\mu})$ operations. Finally, according to the properties of the subspace polynomial $s_{\mu}(x)$, the evaluation $s_{\mu}(\beta)$ or $s_{\mu}(v_{\mu})$ has the same complexity as a field multiplication. Hence, the total computational complexity of Algorithm~\ref{alg:ex_IFFT} is ${O}(\mu2^{\mu})$.
	\end{pf}
\end{lemma}

Since
$n - k = 2^{\mu}$, the complexity of calling Algorithm~\ref{alg:ex_IFFT} twice in Algorithm~\ref{alg:FDMA} is ${O}((n-k)\log(n-k))$.
It follows that the computational complexity of the FDMA is ${O}((n-k)^2)$.

\subsection{Fast Modular Approach}

{In this subsection, we present the FMA for solving~\eqref{eq:key_WB}. The idea behind the FMA is that ${\phi}_i^{j}$ and $\Psi_1^j$ need not to be computed at each iteration until enough information has been collected. Define the notation
$\Psi_j^l = \Psi_l\Psi_{l-1}\cdots\Psi_j$ for any $1 \leq j \leq l \leq 2^{\mu}$. Recall that
\begin{align*}
	\phi_i^j(W(x), N(x)) &= \phi_i \circ \psi_1^j(W(x), N(x)) = \phi_i((W(x), N(x))\Psi_1^j)\\
	&= (W(\omega_{i-1}), N(\omega_{i-1})\Psi_1^j(\omega_{i-1})\begin{pmatrix}
	d_i\\
	g_i
	\end{pmatrix}
\end{align*}
and 
$\Psi_1^j = \Psi_j\Psi_{j - 1}\cdots\Psi_1 = \Psi_{j/2 + 1}^{j}\Psi_{1}^{j/2}$
if $j$ is even. Hence, if we first compute the irreducible basis matrix $\Psi_1^{2^{\mu - 1}}$ in $\ker(\phi_1)\cap\ker(\phi_2)\cap\dots\cap\ker(\phi_{2^{\mu - 1}})$, which is a subproblem of~\eqref{eq:key_WB}, then Algorithm~\ref{FFTX} can be used to obtain $\Psi_{1}^{2^{\mu - 1}}(\omega_{i - 1})$ for $i = 1, 2, \dots, 2^{\mu}$ by setting $\beta = 0$. Next, we  update $\phi_i^{2^{\mu - 1}}$ for $i = 2^{\mu - 1} + 1, 2^{\mu - 1} + 2, \dots, 2^{\mu}$. Given $\phi_i^{2^{\mu - 1}}$ for $i = 2^{\mu - 1} + 1, 2^{\mu - 1} + 2, \dots, 2^{\mu}$, we then compute $\Psi_{2^{\mu - 1} + 1} ^ {2^{\mu}}$, which can be obtained in a similar way. 
Finally, we obtain the product $\Psi_1 ^ {2^\mu} = \Psi_{2^{\mu - 1} + 1} ^ {2^{\mu}}\Psi_1^{2^{\mu - 1}}$ by the well-known convolution theorem. More precisely, if $\Psi_1^{2^{\mu - 1}}(\omega_{i-1}), \Psi_{2^{\mu - 1} + 1} ^ {2^{\mu}}(\omega_{i-1})$ for $i = 1, 2, \dots, 2^{\mu} + 1$ are available,  $\Psi_1 ^ {2^\mu}(\omega_{i-1})$ can be computed by simple matrix multiplication. Then Algorithm~\ref{alg:ex_IFFT} can be used to obtain $\Psi_1 ^ {2^\mu}$. Obviously, $\Psi_{2^{\mu - 1} + 1} ^ {2^{\mu}}(\omega_{i-1})$ can also be obtained using Algorithm~\ref{FFTX}.}

{We can generalize the above idea. In general, if $\omega_{i - 1} = \omega_{i-j} + \omega_{j - 1}$ for $i = j, j + 1, \dots, j + 2^{\mu} - 1$, then we have
$	(\omega_{j - 1}, \omega_{j}, \dots, \omega_{j + 2^{\mu} - 2}) = (\omega_0 + \omega_{j-1}, \omega_{1} + \omega_{j-1}, \dots, \omega_{2^{\mu} - 1} + \omega_{j-1}).$
Thus, Algorithm~\ref{FFTX} can be used for evaluating a polynomial at points $\omega_{j - 1}, \omega_{j}, \dots, \omega_{j + 2^{\mu} - 2}$ by setting $\beta = \omega_{j-1}$. Hence, if we want to obtain $\Psi_{j}^{j + 2^{\mu} - 1}$ with input $\phi_j^{j-1}, \phi_{j+1}^{j-1},\dots, \phi_{j+2^{\mu} - 1}^{j-1}$,  we first compute $\Psi_j^{j + 2^{\mu-1} - 1}$. Then we update $\phi_i^{j + 2^{\mu-1} - 1}$ for $i = j + 2^{\mu - 1}, \dots, j + 2^{\mu} - 1$. Next, based on these updated homomorphisms, we  compute ${\Psi}_{j + 2^{\mu-1}}^{j + 2^{\mu} - 1}$ by induction. Finally, we obtain $\Psi_{j}^{j + 2^{\mu} - 1} = {\Psi}_{j + 2^{\mu-1}}^{j + 2^{\mu} - 1}\Psi_j^{j + 2^{\mu-1} - 1}$. A detailed description of this procedure is given in Algorithm~\ref{alg:FMA}. Note that since $\bar{X}_0(x) = 1$, we use $1$ instead of $\bar{X}_0(x)$ for clarity.}

\begin{breakablealgorithm}
	\caption{Fast Modular Approach (FMA)}
	\label{alg:FMA}
	\begin{algorithmic}[1]
		\REQUIRE
		{$\{\omega_{i-1}, d_i^{j-1}, g_i^{j-1}\}, i = j,  j + 1, \dots, j + 2^{\mu} - 1$, which satisfies $\omega_{i - 1} = \omega_{i-j} + \omega_{j - 1}$ and $r_1^{j - 1}, r_2^{j - 1}$.}
		
		\ENSURE
		$\Psi_{j}^{j + 2^{\mu} - 1}, {r}_1^{j + 2^{\mu} - 1}, {r}_2^{j + 2^{\mu} - 1}$, where the polynomials are represented with respect to $\bar{\X}$;
		
		\IF {$\mu = 0$}
		\IF {${g}_{j}^{j - 1} = 0\ \text{or}\ ({d}_{j}^{j - 1}\neq 0\ \text{and}\ r_1^{j - 1} < r_2^{j - 1})$}
		\STATE
		\begin{equation*}
		\Psi_{j}^{j + 2^{\mu} - 1}=
		\begin{pmatrix}
		-{g}_{j}^{j - 1} & {d}_{j}^{j - 1} \\
		\omega_1\bar{X}_1(x) - \omega_{j - 1} & 0
		\end{pmatrix}
		\end{equation*}
		\STATE ${r}_1^{j + 2^{\mu} - 1} = r_2^{j - 1} , {r}_2^{j + 2^{\mu} - 1} = r_1^{j - 1} + 2$
		\ELSE
		\STATE
		\begin{equation*}
		\Psi_{j}^{j + 2^{\mu} - 1}=
		\begin{pmatrix}
		-{g}_{j}^{j - 1} & {d}_{j}^{j - 1} \\
		0 & \omega_1\bar{X}_1(x) - \omega_{j - 1}
		\end{pmatrix}
		\end{equation*}
		\STATE ${r}_1^{j + 2^{\mu} - 1} = r_1^{j - 1}, {r}_2^{j + 2^{\mu} - 1} = r_2^{j - 1} + 2$
		\ENDIF
		\ELSE
		\STATE
		{Call $\operatorname{FMA}(\{\omega_{i-1}, d_i^{j-1}, g_i^{j-1} \}, i = j, j + 1,\dots, j + 2^{\mu-1} - 1, r_1^{j-1}, r_2^{j-1})$ to obtain} $$(\Psi_j^{j + 2^{\mu-1} - 1}, r_1^{j + 2^{\mu-1} - 1}, r_2^{j + 2^{\mu-1} - 1})$$
		\STATE
		Call Algorithm~\ref{FFTX} to obtain $\Psi_j^{j + 2^{\mu-1} - 1}(\omega_{j - 1}), \dots, \Psi_j^{j + 2^{\mu-1} - 1}(\omega_{j + 2^{\mu} - 2})$
		
		and compute $\Psi_j^{j + 2^{\mu-1} - 1}(\omega_{j + 2^{\mu} - 1})$.
		\FOR {$i = j + 2^{\mu-1}, j + 2^{\mu-1} + 1, \dots, j + 2^{\mu} - 1$}
		\STATE \begin{equation*}
		\begin{pmatrix}
		{d}_i^{j + 2^{\mu-1} - 1} \\
		{g}_i^{j + 2^{\mu-1} - 1}
		\end{pmatrix}
		 = 
		\Psi_j^{j + 2^{\mu-1} - 1}(\omega_{i - 1})
		\begin{pmatrix}
		d_{i}^{j - 1} \\
		g_{i}^{j - 1}
		\end{pmatrix}.
		\end{equation*}
		\ENDFOR
		\STATE
		 Let $l = j + 2^{\mu-1}$.
		\STATE 
		{Call $\operatorname{FMA}(\{\omega_{h - 1}, d_h^{l-1}, g_h^{l-1} \}, h = l, \dots, l + 2^{\mu - 1} - 1 , r_1^{l - 1}, r_2^{l - 1})$ to obtain} $$({\Psi}_{l}^{l + 2^{\mu - 1} - 1}, {r}_1^{l + 2^{\mu - 1} - 1}, {r}_2^{l + 2^{\mu - 1} - 1})$$
		\STATE
		Call Algorithm~\ref{FFTX} to obtain ${\Psi}_{l}^{l + 2^{\mu - 1} - 1}(\omega_{j - 1}), \dots, {\Psi}_{l}^{l + 2^{\mu - 1} - 1}(\omega_{j + 2^{\mu} - 2})$
		
		and compute ${\Psi}_{l}^{l + 2^{\mu - 1} - 1}(\omega_{j + 2^{\mu} - 1})$.
		\FOR {$i = j,  j + 1, \dots, j + 2^{\mu}$} 
		\STATE$
		\Psi_{j}^{j + 2^{\mu} - 1}(\omega_{i-1}) = {\Psi}_{l}^{l + 2^{\mu - 1} - 1}(\omega_{i-1}) \Psi_j^{j + 2^{\mu-1} - 1}(\omega_{i-1})$.
		\ENDFOR
		\STATE
		Call Algorithm~\ref{alg:ex_IFFT} to obtain each component of $\Psi_{j}^{j + 2^{\mu} - 1}$.
		\ENDIF
		
		\RETURN $\Psi_{j}^{j + 2^{\mu} - 1}, {r}_1^{j + 2^{\mu} - 1}, {r}_2^{j + 2^{\mu} - 1}$.
	\end{algorithmic}
\end{breakablealgorithm}

\begin{lemma}\label{lem:FMA}
	 {Given the input $\{\omega_{i-1}, d_i^{j-1}, g_i^{j-1}\}, i = j,  j + 1, \dots, j + 2^{\mu} - 1$, which satisfies $\omega_{i - 1} = \omega_{i-j} + \omega_{j - 1}$ and $r_1^{j - 1}, r_2^{j - 1}$,  Algorithm~\ref{alg:FMA} outputs $\Psi_{j}^{j + 2^{\mu} - 1}$, ${r}_1^{j + 2^{\mu} - 1}$, ${r}_2^{j + 2^{\mu} - 1}$.}
	
	\begin{pf}
		The proof is by induction on $\mu$.
		If $\mu = 0$, since we have $x = \omega_1\bar{X}_1(x)$ and $\bar{X}_0(x) = 1$, then according to the proof of Lemma~\ref{lem:basis_mat}, Algorithm~\ref{alg:FMA} outputs the desired answer $\Psi_j^j = \Psi_j$ , $r_1^j$, $r_2^j$ for any $j$. Therefore, the claim holds for $\mu = 0$.
		
		Suppose that the claim holds for $0, 1, \dots, \mu - 1$. Then $\Psi_j^{j + 2^{\mu-1} - 1}$,  $r_1^{j + 2^{\mu-1} - 1}$, $r_2^{j + 2^{\mu-1} - 1}$ can be obtained by the recursive call of the FMA in line 10 by induction. Since $\omega_{i - 1} = \omega_{i-j} + \omega_{j - 1}$, it follows that $
			(\omega_{j-1}, \omega_{j}, \dots,\omega_{j + 2^\mu - 2}) = (\omega_{0} + \omega_{j-1}, \omega_{1} + \omega_{j-1}, \dots, \omega_{2^\mu - 1} + \omega_{j - 1})$.
		Therefore, Algorithm~\ref{FFTX} can be called for evaluating the matrix $\Psi_j^{j + 2^{\mu-1} - 1}$ at points $\omega_{j-1}, \dots, \omega_{j + 2^\mu - 2}$ by setting $\tau = \mu$ and $\beta = \omega_{j-1}$. The evaluation $\Psi_j^{j + 2^{\mu-1} - 1}(\omega_{j + 2^{\mu} - 1})$ can be computed immediately. Because $\Psi_1^{j + 2^{\mu-1} - 1}(\omega_{i - 1}) = \Psi_j^{j + 2^{\mu-1} - 1}(\omega_{i - 1})\Psi_1^{j - 1}(\omega_{i - 1})$, we have
		\begin{equation*}
		\begin{pmatrix}
		{d}_i^{j + 2^{\mu-1} - 1} \\
		{g}_i^{j + 2^{\mu-1} - 1}
		\end{pmatrix}
		 = 
		\Psi_1^{j + 2^{\mu-1} - 1}(\omega_{i - 1})
		\begin{pmatrix}
		d_i \\
		g_i
		\end{pmatrix}
		=
		\Psi_j^{j + 2^{\mu-1} - 1}(\omega_{i - 1})
		\begin{pmatrix}
		d_{i}^{j - 1} \\
		g_{i}^{j - 1}
		\end{pmatrix}
		\end{equation*}
		for $i = j + 2^{\mu-1}, j + 2^{\mu-1} + 1, \dots, j + 2^{\mu} - 1$.
		Let $l = j + 2^{\mu-1}$. 
		For $h = l, l + 1, \dots, l + 2^{\mu-1} - 1$, we have
		$
			\omega_{h-1} = \omega_{h-j} + \omega_{j-1}
			= \omega_{h-l} + \omega_{l - j} + \omega_{j-1}
			= \omega_{h-l} + \omega_{l - 1}, 
		$
		where the first and the last equalities hold by induction and the second equality is true because $h - l < 2^{\mu-1}$ and $\omega_{l-j} = \omega_{2^{\mu - 1}}$. Hence, the recursive call of the FMA in line 16 outputs the desired ${\Psi}_{l}^{l + 2^{\mu - 1} - 1}$, ${r}_1^{l}$, ${r}_2^{l + 2^{\mu - 1} - 1}$ by induction.
		
		Next, we show that the degrees of the components of $\Psi_{j}^{j + 2^{\mu} - 1}$ are less than or equal to $2^{\mu}$, which implies that $\Psi_{j}^{j + 2^{\mu} - 1}$ is determined uniquely by $\Psi_{j}^{j + 2^{\mu} - 1}(\omega_{i-1}), i = j, j + 1, \dots, j+ 2^\mu$. It is clear that this conclusion is true for $\mu = 0$. Suppose that it is also true for $1, 2, \dots, \mu-1$. Then the degrees of the components of $\Psi_{j}^{j + 2^{\mu} - 1}$ must be less than or equal to $2^{\mu}$, since the degrees of the components of ${\Psi}_{l}^{l + 2^{\mu - 1} - 1}$ and $\Psi_j^{j + 2^{\mu-1} - 1}$ are less than or equal to $2^{\mu-1}$ by induction. Hence, we can determine $\Psi_{j}^{j + 2^{\mu} - 1}$ by Algorithm~\ref{alg:ex_IFFT} once $\Psi_{j}^{j + 2^{\mu} - 1}(\omega_{i-1}), i = j, j + 1, \dots, j+ 2^\mu$, have been obtained. This completes the proof.
	\end{pf}
\end{lemma}

	\begin{theorem}
		{Given $\{\omega_{i-1}, d_i^0, g_i^0 \}, i = 1, 2, \dots, 2^{\mu}, r_1^{0}$, $r_2^{0}$, Algorithm~\ref{alg:FMA} outputs $\Psi_{1}^{2^{\mu}}$, ${r}_1^{2^{\mu}}$, ${r}_2^{2^{\mu}}$.}
		\begin{pf}
			Since $\omega_{0}$ is the additive identity in $\F_{2^m}$, we have $\omega_{i-1} = \omega_{i-j} + \omega_{j-1}$ for $i = 1, 2, \dots,  2^{\mu}$ and $j = 1$. The theorem then follows by Lemma~\ref{lem:FMA}. 
		\end{pf}
	\end{theorem}
	{For solving \eqref{eq:key_WB}, we set $d_i^0 = u(\omega_{i-1}), g_i^0 = 1$ for $i = 1, 2, \dots, 2^{\mu}$, $r_1^0 = 0, r_2^0 = 1$.}
	
	{We now analyze the computational complexity of Algorithm~\ref{alg:FMA}.} Denote the complexity of Algorithm~\ref{alg:FMA} by $T(2^\mu)$. If $\mu = 0$, the algorithm outputs the solution in a straightforward manner with complexity ${O}(1)$. Assume that  $\mu > 1$. Two recursive calls take $2T(2^{\mu - 1})$. The complexity of calling Algorithm~\ref{FFTX} twice for evaluating ${\Psi}_{l}^{l + 2^{\mu - 1} - 1}$ and $\Psi_j^{j + 2^{\mu-1} - 1}$ at points $\omega_{j-1}, \dots, \omega_{j + 2^{\mu} - 2}$ is ${O}(\mu2^\mu)$, and the complexity of evaluating ${\Psi}_{l}^{l + 2^{\mu - 1} - 1}$ and $\Psi_j^{j + 2^{\mu-1} - 1}$ at a single point $\omega_{j + 2^{\mu} - 1}$ is ${O}(2^\mu)$. In addition, computing $\phi_i^{j + 2^{\mu-1} - 1}$ for $i = j + 2^{\mu-1}, \dots, j+2^{\mu}-1$ involves ${O}(2^{\mu-1})$ operations, while the matrix multiplication between ${\Psi}_{l}^{l + 2^{\mu - 1} - 1}$ and $\Psi_j^{j + 2^{\mu-1} - 1}$ in the frequency domain involves ${O}(2^{\mu})$ operations. Finally, the complexity of calling Algorithm  \ref{alg:ex_IFFT} four times is ${O}(\mu2^{\mu})$. It follows that $
		T(2^\mu) = 2T(2^{\mu - 1}) + {O}(\mu2^{\mu})$ and $T(2^\mu) = {O}(2^\mu\log^2(2^\mu))$.
	As $n-k = 2^{\mu}$, we have $T(n-k) = {O}((n-k)\log^2(n-k))$.
	
	Evidently, the computational complexity of this new decoding algorithm is $O(n\log(n-k) + (n-k)\log^2(n-k))$. In the next section, we show that the FMA has a smaller constant factor than the Half-GCD algorithm proposed in~\cite{FFT2016}. This implies that the new algorithm has the smallest constant factor to  date. It should be mentioned that although the complexity of the FDMA is $O((n-k)^2)$, it is more efficient for decoding short codes, which we shall see in the next section.
	
{
The complete decoding algorithm is presented in Algorithm \ref{alg:decoding}. Note that this method can be generalized to arbitrary code length $n$ and code dimension $k$ and its complexity remains to be $O(n\log(n-k) + (n-k)\log^2(n-k))$. Detailed discussion is provided in Appendix \ref{app:arbitrary}.
\begin{algorithm}[h]
	\caption{{Decoding Algorithm}}
	\label{alg:decoding}
	\begin{algorithmic}[1]
		\REQUIRE
		Received vector $\mathbf{r} = \mathbf{F} + \mathbf{e}$.
		\ENSURE
		The codeword $\mathbf{F}$.
		\STATE Compute the syndrome polynomial $u(x)$ according to \eqref{eq:synd}.
		\STATE Evaluate $u(x)$ at points $\omega_0, \omega_1, \dots, \omega_{2^{\mu}-1}$ by Algorithm \ref{FFTX}.
		\STATE Given $\phi_i(W(x), N(x)) = u(\omega_{i-1})W(\omega_{i-1}) + N(\omega_{i-1}), i = 1, 2, \dots, 2^{\mu}$, compute the error locator polynomial $\lambda(x)$ and the error evaluator polynomial $z(x)$ by Algorithms \ref{alg:FDMA} or \ref{alg:FMA}.
		\STATE Find the error locations by \eqref{eq:FFT_search}.
		\STATE Compute the error pattern $\mathbf{e}$ by \eqref{eq:Forney}.
		\RETURN
		$\mathbf{r} + \mathbf{e}$.
	\end{algorithmic}
\end{algorithm}

\section{Comparison and Analysis}\label{sec:comparison}

In this section, we compare the proposed algorithm with other methods.

\subsection{Comparison with Conventional Syndrome-Based Decoding}
The most commonly used decoding algorithm for RS codes is  syndrome-based decoding, which is based on  Horner's rule. 
Here, we compare the algorithm described in Section~\ref{sec:newdecoding} with  syndrome-based decoding. Three RS codes, namely, the $(255, 223)$ RS code over $\F_{2^8}$, the $(1023, 895)$ RS code over $\F_{2^{10}}$, and the $(4095, 3583)$ RS code over $\F_{2^{12}}$, are selected, and the comparisons are done by counting the numbers of field multiplications, additions, and divisions required by the two decoding algorithms. 
Since there is no field inversion in the FDMA or FMA,  a modified BM algorithm that involves no field inversion, called the reformulated inversionless BM algorithm (RiBM),   is chosen for comparison. Further discussion of the RiBM algorithm can be found in~\cite{Sarw2001}. Tables~\ref{tab:RS256}, \ref{tab:RS1024}, and \ref{tab:RS4096} present the comparisons in detail. {According to these tables, the proposed algorithm saves 68\%, 74\%, and 90\% multiplications and 60\%, 74\%, and 84\% additions over $\F_{2^8}$, $\F_{2^{10}}$, and $\F_{2^{12}}$, respectively.} Evidently, the proposed algorithm is $10$ times faster than conventional decoding on a given machine when decoding $(4095, 3583)$ RS codes.
Note that the FMA is suitable for RS codes with a medium or long length, while the FDMA is more efficient when decoding short RS codes.\footnote{The reason that the FMA performs worse than the FDMA for short codes is due to the hidden cost for dividing the problem and merging the solutions obtained from the subproblems when performing the divide and conquer approach (FMA).}  It can be seen from Table~\ref{tab:RS4096} that the complexity of the FMA is significantly better than that of the RiBM algorithm for medium or long RS codes.

\begin{table}[h]
\caption{{Complexity Comparison Between Syndrome-Based Decoding (RiBM) for the $(255,223)$ RS Code and the New Decoding (FDMA) for the $(256,224)$ RS Code over $\F_{2^8}$}}
\begin{center}
	\begin{tabular}{|c|c|c|c|c|c|c|}
		\hline
		\multirow{2}{*}{Components} & \multicolumn{3}{c|}{{Syndrome-based decoding (RiBM)}} & \multicolumn{3}{c|}{{New decoding (FDMA)}}\\
		\cline{2-7}
		&Mul. & Add. & Div. & Mul. & Add. & Div.\\
		\hline
		Syndrome& 8,160 & 8,160 & 0 & 752 & 1,696 & 0\\
		\hline
		{Key equation} & 3,136 & 1,568 & 0 & 3,233 & 2,244 & 0\\
		\hline
		Chien search& 4,335 & 4,335 & 0 & 640 & 1,280 & 0\\
		\hline
		Formal derivative & 0 & 0 & 0 & 80 & 80 & 0\\
		\hline
		Forney's formula& 544 & 528 & 16 & 544 & 528 & 16\\
		\hline
		{Total} & {16,175} & {14,591} & 16 & {5,249} & {5,828} & 16\\
		\hline
	\end{tabular}
\end{center}
\label{tab:RS256}
\end{table}

\begin{table}[h]
	\caption{{Complexity Comparison Between Syndrome-Based Decoding (RiBM) for the $(1023,895)$ RS Code and the New Decoding (FDMA) for the $(1024, 896)$ RS Code over $\F_{2^{10}}$}}
	\begin{center}
		\begin{tabular}{|c|c|c|c|c|c|c|}
			\hline
			\multirow{2}{*}{Components} & \multicolumn{3}{c|}{{Syndrome-based decoding (RiBM)}} & \multicolumn{3}{c|}{{New decoding (FDMA)}}\\
			\cline{2-7}
			&Mul. & Add. & Div. & Mul. & Add. & Div.\\
			\hline
			Syndrome& 130,944 & 130,944 & 0 & 4,160 & 9,088 & 0\\
			\hline
			{Key equation} & 49,408 & 24,704 & 0 & 49,921 & 33,796 & 0\\
			\hline
			Chien search& 66,495 & 66,495 & 0 & 3,584 & 7,168 & 0\\
			\hline
			Formal derivative & 0 & 0 & 0 & 448 & 448 & 0\\
			\hline
			Forney's formula& 8,320 & 8,256 & 64 & 8,320 & 8,256 & 64\\
			\hline
			{Total} & {255,167} & {230,399} & 64 & {66,433} & {58,756} & 64\\
			\hline
		\end{tabular}
	\end{center}
	\label{tab:RS1024}
\end{table}

\begin{table}[h]
	\caption{{Complexity Comparison Between Syndrome-Based Decoding (RiBM) for the $(4095,3583)$ RS Code and the New Decoding (FMA) for the $(4096, 3584)$ RS Code over $\F_{2^{12}}$}}
	\begin{center}
		\begin{tabular}{|c|c|c|c|c|c|c|}
			\hline
			\multirow{2}{*}{Components} & \multicolumn{3}{c|}{{Syndrome-based decoding (RiBM)}} & \multicolumn{3}{c|}{{New decoding (FMA)}}\\
			\cline{2-7}
			&Mul. & Add. & Div. & Mul. & Add. & Div.\\
			\hline
			Syndrome& 2,096,640 & 2,096,640 & 0 & 21,248 & 45,568 & 0\\
			\hline
			{Key equation} & 787,456 & 393,728 & 0 & 239,616 & 357,372 & 0\\
			\hline
			Chien search& 1,052,415 & 1,052,415 & 0 & 18,432 & 36,864 & 0\\
			\hline
			Formal derivative & 0 & 0 & 0 & 2,304 & 2,304 & 0\\
			\hline
			Forney's formula& 131,584 & 131,328 & 256 & 131,584 & 131,328 & 256\\
			\hline
			{Total} & {4,068,095} & {3,674,111} & 256 & {413,184} & {573,436} & 256\\
			\hline
		\end{tabular}
	\end{center}
	\label{tab:RS4096}
\end{table}

\subsection{Comparison with Other RS Algorithms Based on FFT}
There are many other efficient RS algorithms based on various FFT methods~\cite{lin2007fast,wu2012reduced,bellini2011structure,fedorenko2019efficient}. Note that it has been shown in~\cite{Lin20161} that the additive FFT based on the Taylor expansion is worse than the FFT used here in terms of additive complexity. Thus, we do not consider the algorithm in~\cite{gao2010additive} for comparison.
Table \ref{tab:FFT_comp} compares the new decoding algorithm with the methods in~\cite{lin2007fast,wu2012reduced,bellini2011structure,fedorenko2019efficient} by counting the field operations in the syndrome computation. The result shows that the new algorithm has the lowest additive complexity and a medium multiplicative complexity. It should be mentioned that although some existing algorithms have a lower multiplicative complexity, they sacrifice a regular structure, which is vital in hardware implementation. 
The new decoding algorithm uses a FFT algorithm in which a butterfly structure is present; see~\cite{FFT2016} for details. This makes the new algorithm suitable for hardware implementation. 
Furthermore, the existing decoding algorithms have no fast key equation solver. Hence, the new algorithm is significantly better than them for decoding medium or long RS codes, as we have seen in Table~\ref{tab:RS4096}.

\begin{table}[ht]
	\caption{Complexity of Syndrome Computation for RS Codes over $\F_{2^m}$}
	\begin{center}
		\begin{tabular}{|c|c|c|c|c|c|c|c|c|c|c|c|}
			\hline
			\multirow{2}{*}{Field} & \multirow{2}{*}{Code} & \multicolumn{2}{c|}{Method in~\cite{lin2007fast}} & \multicolumn{2}{c|}{Method in~\cite{wu2012reduced}} & \multicolumn{2}{c|}{Method in~\cite{bellini2011structure}} & \multicolumn{2}{c|}{Method in~\cite{fedorenko2019efficient}} & \multicolumn{2}{c|}{Proposed algorithm}\\
			\cline{3-12}
			& & Mul. & Add. & Mul. & Add. & Mul. & Add. & Mul. & Add. & Mul. & Add.\\
			\hline
			$\F_{2^8}$ & $(255, 223)$ & 3,060 & 4,998 & 252 & 3,064 & 149 & 2,931 & 138 & 3,064 & 752 & 1,696\\
			\hline
			$\F_{2^{10}}$ &$(1023, 895)$ & 33,620 & 73,185 & 2,868 & 19,339 & 824 & 36,981 &  / & / & 4,160 & 9,088\\
			\hline
		\end{tabular}
	\end{center}
	\label{tab:FFT_comp}
\end{table}

\subsection{Comparison with the Half-GCD Algorithm and the Guruswami-Sudan Algorithm}

The Half-GCD algorithm, proposed in~\cite{FFT2016}, is able to solve~\eqref{eq:key_WB} with complexity ${O}((n-k)\log^{2}(n-k))$. Although the FMA algorithm has the same complexity order as the Half-GCD algorithm, it has a smaller constant factor and a regular structure. It is clear that Algorithm~\ref{alg:FMA} involves two recursive calls, matrix multiplications, eight times $2^{\mu}$-point $\FFT_{\bar{\X}}$, and four times Algorithm~\ref{alg:ex_IFFT}. As a $2^{\mu}$-point $\FFT_{\bar{\X}}$ involves $\frac{1}{2}{\mu}2^{\mu}$  multiplications and ${\mu}2^{\mu}$ additions, if we assume that the multiplication and addition have the same complexity, the constant factor of $\FFT_{\bar{\X}}$ is $1.5$. In other words, a $2^{\mu}$-point $\FFT_{\bar{\X}}$ costs $1.5\mu2^{\mu}$ field operations. Furthermore, the constant factor of Algorithm~\ref{alg:ex_IFFT} is also $1.5$.
Hence, as the matrix multiplications involve ${O}(2^{\mu})$ operations, we have
\begin{align*}
T(2^{\mu}) = 2T(2^{\mu}/2) + (8 + 4) \times 1.5\mu2^{\mu} + o(2^{\mu}\log(2^{\mu}))
< 2T(2^{\mu}/2) + 19\mu2^{\mu}.
\end{align*}
This implies that the constant factor of Algorithm~\ref{alg:FMA} is less than $9.5$. For comparison, the Half-GCD algorithm involves two recursive calls, at least 15 times $2^{\mu}$-point $\FFT_{\bar{\X}}$, and 15 times $2^{\mu}$-point $\IFFT_{\bar{\X}}$. This means that the constant factor of  Half-GCD is at least 22.5. Hence, the FMA has a significantly improved decoding complexity compared with  Half-GCD. Moreover, with some effort, one can show that the FMA does not require that $n-k$ be a power of two. Therefore, the FMA is more flexible than  Half-GCD in real applications.

{Let the list size $\kappa = 1$ and the multiplicity $\upsilon = 1$. The GS algorithm is equivalent to bounded distance decoding. By taking the fast polynomial multiplication into account, fast interpolation algorithms were proposed in \cite{Alekhnovich2005linear, beelen2010key} and \cite{chowdhury2015faster} for solving the key equation of the GS algorithm. Their complexities are $O(n\log^2 n\log\log n)$. Compared with these interpolation algorithms, the method proposed here is faster since there is no factor $\log\log n$.}

\section{Conclusion}\label{sec:conclusion}
We have presented the MA, which is an efficient algorithm for solving the  WB key equation. Based on the MA, a new decoding algorithm for RS codes has been proposed that has the best asymptotic computational complexity and  the smallest constant factor achieved to date. The results of comparisons show that the new decoding algorithm is significantly better than {the} existing methods in terms of complexity when decoding practical RS codes. Since the complexity of the new algorithm is $O(n\log(n-k) + (n-k)\log^2(n-k))$, this makes it possible to use long RS codes in real applications. 
One potential route for future work is to transfer this new algorithm into a circuit design.
{Another interesting issue is to devise a fast list decoding algorithm based on the techniques presented here.} {Finally, whether the proposed algorithm can be used in the one-pass Chase decoding presented in \cite{shany2022a} is open yet.}

\begin{appendices}

\section{}\label{app:arbitrary}

In this section, we first demonstrates that the FFT/IFFT transforms exist for arbitrary $\epsilon$-points and their complexities are $O(\epsilon\log \epsilon)$. Then we present the encoding and decoding algorithm for arbitrary $(n,k)$ RS codes. Let $f(x) = \sum_{l = 0}^{\epsilon-1}\bar{f}_l\bar{X}_l(x)$ and  $\mu$ be the smallest integer such that $2^{\mu}\geq \epsilon$. Obviously, we have $\epsilon > 2^{\mu-1}$.

\subsection{$\epsilon$-points FFT transform}

The $\epsilon$-points FFT transform evaluates $f(x)$ at points $\omega_{0} + \beta, \omega_{1} + \beta, \dots, \omega_{\epsilon - 1} + \beta$. Since $f(x) = \sum_{l = 0}^{2^{\mu} - 1}\bar{f}_l\bar{X}_l(x)$ where $\bar{f}_{\epsilon}, \dots, \bar{f}_{2^{\mu} - 1} = 0$, then Algorithm \ref{FFTX} can be taken to evaluate $f(x)$ at points $\omega_{0} + \beta, \omega_{1} + \beta, \dots, \omega_{2^{\mu} - 1} + \beta$. Furthermore, as $\epsilon > 2^{\mu-1}$, then $2\epsilon > 2^{\mu}$. Thus, the complexity of the $\epsilon$-points FFT transform is at most $O(2^{\mu}\log2^{\mu}) = O(\epsilon\log\epsilon)$.

\subsection{$\epsilon$-points IFFT transform}

Given $f(\omega_0 + \beta), f(\omega_1 + \beta), \dots, f(\omega_{\epsilon - 1} + \beta)$, the $\epsilon$-points IFFT transform computes the coordinate vector of $f(x)$ with respect to $\bar{\X}$, where $\epsilon \geq 1$ and $\deg(f(x)) < \epsilon$. An algorithm for this task has been presented in \cite[Algorithm 3]{Fast2020}. However, for this paper to be self-contained, we again present this algorithm here.

For the $\epsilon$-points IFFT transform where $\epsilon \leq 2^{\mu}$, only $f(\omega_0 + \beta), f(\omega_1 + \beta), \dots, f(\omega_{\epsilon - 1} + \beta)$ are given. Hence, Algorithm \ref{IFFTX} fails to accomplish such a transform. However, Lemma 1 given in~\cite{Fast2020} provides a tool for deriving the $\epsilon$-points IFFT transform. Hereafter, we write
\begin{align*}
	\mathbf{F} &= (F_0, F_1,\dots,F_{2^{\mu} - 1}) = (f(\omega_0 + \beta), f(\omega_1 + \beta), \dots, f(\omega_{2^{\mu} - 1} + \beta)),
\end{align*}
and write the coordinate vector of $f(x)$ with respect to $\bar{\X}$ as
\begin{equation*}
	\bar{\mathbf{f}} = (\bar{f}_0, \bar{f}_1, \dots, \bar{f}_{2^{\mu} - 1}).
\end{equation*}
Furthermore, for any integer $0\leq \gamma \leq \mu$, we write
\begin{equation*}
	\mathbf{F}_{l,\gamma} = (F_{l\cdot 2^{\gamma}}, F_{l\cdot 2^{\gamma} + 1}, \dots, F_{(l + 1)\cdot 2^{\gamma} - 1}), \bar{\mathbf{f}}_{l, \gamma} = (\bar{f}_{l\cdot 2^{\gamma}}, \bar{f}_{l\cdot 2^{\gamma} + 1}, \dots, \bar{f}_{(l + 1)\cdot 2^{\gamma} - 1})
\end{equation*}
for $0\le \ell\le 2^{\mu-\gamma}-1.$
\begin{lemma}[\cite{Fast2020}, Lemma 1]\label{lem:IFFT_property}
	For any integer $0\leq \gamma \leq \mu$, we have
	\begin{equation*}
		\sum_{l = 0}^{2^{\mu-\gamma} - 1}\IFFT_{\bar{\mathbb{X}}}(\mathbf{F}_{l,\gamma}, \gamma, \omega_{l\cdot 2^{\gamma}} + \beta) = \bar{\mathbf{f}}_{2^{\mu-\gamma} - 1, \gamma}.
	\end{equation*}
\end{lemma}

Based on the above lemma, Algorithm \ref{alg:IFFTX_epsilon} presents the $\epsilon$-points IFFT transform. Note that we remove the constraint $\epsilon > 2^{\mu - 1}$ in Algorithm \ref{alg:IFFTX_epsilon}.

\begin{breakablealgorithm}
	\caption{$\epsilon$-points IFFT Transform \cite[Algorithm 3]{Fast2020}}
	\label{alg:IFFTX_epsilon}
	\begin{algorithmic}[1]
		\REQUIRE
		$\{(F_0, F_1, \dots, F_{\epsilon - 1}), \epsilon, \mu, \beta\}$, where $\epsilon \leq 2^{\mu}$.
		\ENSURE
		$\{\bar{\mathbf{f}}, \mathbf{F}\}$
		such that $\bar{f}_{\epsilon}, \dots, \bar{f}_{2^{\mu} - 1} = 0$ and
		$\FFT_{\bar{\mathbb{X}}}(\bar{\mathbf{f}}, \mu, \beta) = \mathbf{F}$.
		
		\IF {$\mu = 0$} \RETURN $\{\bar{\mathbf{f}} = (F_0), \mathbf{F} = (F_0)\}$.
		\ENDIF
		\IF {$\epsilon \leq 2^{\mu - 1}$}
			\STATE Call Algorithm \ref{alg:IFFTX_epsilon} with input 
			$\{(F_0, F_1, \dots, F_{\epsilon - 1}), \epsilon, \mu - 1, \beta\}$
			 to obtain $\{\bar{\mathbf{f}}_{0, \mu - 1}, \mathbf{F}_{0, \mu - 1}\}$.
			 \STATE Call Algorithm \ref{FFTX} to obtain $\mathbf{F}_{1, \mu - 1} = \FFT_{\bar{\mathbb{X}}}(\bar{\mathbf{f}}_{0, \mu - 1},\mu - 1, \omega_{2^{\mu - 1}} +\beta)$.
			 \STATE Set $\bar{\mathbf{f}}_{1, \mu - 1} = \mathbf{0}$.
		\ELSE
			\STATE Call Algorithm \ref{IFFTX} to obtain $\mathbf{w} = \IFFT_{\bar{\mathbb{X}}}(\mathbf{F}_{0, \mu - 1}, \mu - 1, \beta)$.
			\STATE Call Algorithm \ref{FFTX} to obtain $\mathbf{w}^{'} = \FFT_{\bar{\mathbb{X}}}(\mathbf{w}, \mu - 1, \omega_{2^{\mu - 1}} + \beta)$.
			\STATE Let the input be
			$\{(F_{2^{\mu-1}}, \dots, F_{\epsilon - 1}) + (w_0^{'}, \dots, w_{\epsilon - 1 - 2^{\mu-1} }^{'}), \epsilon - 2^{\mu - 1}, \mu - 1, \omega_{2^{\mu - 1}} + \beta \}$.
			Call Algorithm \ref{alg:IFFTX_epsilon} to get $\{\bar{\mathbf{f}}_{1, \mu - 1}, \mathbf{F}_{1, \mu - 1}^{'}\}$.
			\STATE Set $\bar{\mathbf{f}}_{0, \mu - 1} = \mathbf{w} + \frac{s_{\mu - 1}(\beta)}{s_{\mu - 1}(v_{\mu - 1})}\bar{\mathbf{f}}_{1, \mu - 1}$ and $\mathbf{F}_{1,\mu - 1} = \mathbf{F}_{1, \mu - 1}^{'} + \mathbf{w}^{'}$.
		\ENDIF
		\RETURN $\{\bar{\mathbf{f}}, \mathbf{F}\}$.
	\end{algorithmic}
\end{breakablealgorithm}

 The following we present a lemma that has been mentioned in \cite{Fast2020} and provide a more comprehensive proof here.
 
\begin{lemma}[\cite{Fast2020}, Lemma 2]
	Algorithm \ref{alg:IFFTX_epsilon} outputs $\{\bar{\mathbf{f}}, \mathbf{F}\}$ such that $\FFT_{\bar{\mathbb{X}}}(\bar{\mathbf{f}}, \mu, \beta) = \mathbf{F}$, and its complexity is $O(\epsilon\log\epsilon)$.
	
	\begin{pf}
		We prove this conclusion by induction on $\mu$.
		
		If $\mu = 0$, then $\epsilon = 1$. It is easy to check that $\{\bar{\mathbf{f}} = (F_0), \mathbf{F} = (F_0)\}$ is the desired result.
		
		Now assume that, for $0, 1, \dots, \mu - 1$, Algorithm \ref{alg:IFFTX_epsilon} outputs the desired solution. If $\epsilon \leq 2^{\mu - 1}$, then by induction, calling Algorithm \ref{alg:IFFTX_epsilon} with input $\{(F_0, F_1, \dots, F_{\epsilon - 1}), \epsilon, \mu - 1, \beta\}$ outputs $\{\bar{\mathbf{f}}_{0, \mu - 1}, \mathbf{F}_{0, \mu - 1}\}$ which satisfies $\FFT_{\bar{\mathbb{X}}}(\bar{\mathbf{f}}_{0, \mu - 1}, \mu - 1, \beta) = \mathbf{F}_{0, \mu - 1}$. Furthermore, as $\bar{f}_{\epsilon}, \dots, \bar{f}_{2^{\mu} - 1} = 0$, then the transform $\FFT_{\bar{\mathbb{X}}}(\bar{\mathbf{f}}_{0, \mu - 1},\mu - 1, \omega_{2^{\mu - 1}} +\beta) = \mathbf{F}_{1, \mu - 1}$. The claim holds.
		
		On the other hand, if $\epsilon > 2^{\mu - 1}$, according to Lemma \ref{lem:IFFT_property}, we have
		\begin{equation*}
			\IFFT_{\bar{\mathbb{X}}}(\mathbf{F}_{0, \mu - 1}, \mu - 1, \beta) + \IFFT_{\bar{\mathbb{X}}}(\mathbf{F}_{1, \mu - 1}, \mu - 1, \omega_{2^{\mu-1}} + \beta) = \bar{\mathbf{f}}_{1, \mu - 1}.
		\end{equation*}
		Let $\mathbf{w} = \IFFT_{\bar{\mathbb{X}}}(\mathbf{F}_{0, \mu - 1}, \mu - 1, \beta)$ and $\mathbf{w}^{'} = \FFT_{\bar{\mathbb{X}}}(\mathbf{w}, \mu - 1,\omega_{2^{\mu - 1}} + \beta)$. It follows that
		\begin{align}
			&\mathbf{w} + \IFFT_{\bar{\mathbb{X}}}(\mathbf{F}_{1, \mu - 1}, \mu - 1, \omega_{2^{\mu-1}} + \beta) = \bar{\mathbf{f}}_{1, \mu - 1},\notag\\
			\Rightarrow &\IFFT_{\bar{\mathbb{X}}}(\mathbf{F}_{1, \mu - 1} + \mathbf{w}^{'}, \mu - 1, \omega_{2^{\mu-1}} + \beta) = \bar{\mathbf{f}}_{1, \mu - 1},\notag\\
			\Rightarrow &\FFT_{\bar{\mathbb{X}}}(\bar{\mathbf{f}}_{1, \mu - 1}, \mu - 1, \omega_{2^{\mu-1}} + \beta) = \mathbf{F}_{1, \mu - 1} + \mathbf{w}^{'},\label{eq:IFFT_recursive}
		\end{align}
		where the second identity holds since the transform $\IFFT_{\bar{\mathbb{X}}}$ is linear. Note that since $\mathbf{F}_{0, \mu - 1}$ is known, then $\mathbf{w}$ and $\mathbf{w}^{'}$ are also known. As $\bar{f}_{2^{\mu - 1}}, \dots, \bar{f}_{\epsilon - 1}$, $F_{2^{\mu - 1}}, \dots, F_{\epsilon - 1}$ are known and $\bar{f}_{\epsilon}, \dots, \bar{f}_{2^{\mu} - 1} = 0$, the equation \eqref{eq:IFFT_recursive} can be solved by Algorithm \ref{alg:IFFTX_epsilon} inductively such that $\FFT_{\bar{\mathbb{X}}}(\bar{\mathbf{f}}_{1, \mu - 1}, \mu - 1, \omega_{2^{\mu-1}} + \beta) = \mathbf{F}_{1, \mu - 1}^{'} = \mathbf{F}_{1, \mu - 1} + \mathbf{w}^{'}$. Furthermore, one has $\mathbf{F}_{1, \mu - 1} = \mathbf{F}_{1, \mu - 1}^{'} + \mathbf{w}^{'}$. According to the recursive structure of Algorithm \ref{FFTX}, one has $\mathbf{w} = \bar{\mathbf{f}}_{0, \mu - 1} + \frac{s_{\mu - 1}(\beta)}{s_{\mu - 1}(v_{\mu - 1})}\bar{\mathbf{f}}_{1, \mu - 1}$. Then we obtain that $\bar{\mathbf{f}}_{0, \mu - 1} = \mathbf{w} + \frac{s_{\mu - 1}(\beta)}{s_{\mu - 1}(v_{\mu - 1})}\bar{\mathbf{f}}_{1, \mu - 1}$. Now we have proven the correctness of Algorithm \ref{alg:IFFTX_epsilon}.
		
		It remains to discuss the complexity. Denote the complexity of Algorithm \ref{alg:IFFTX_epsilon} by $T(2^{\mu})$. If $\mu = 0$, the complexity is $O(1)$. Now assume that $\mu > 1$. If $\epsilon \leq 2^{\mu - 1}$,  we have $T(2^{\mu}) = T(2^{\mu - 1}) + O(2^{\mu}\log2^{\mu})$. Conversely, if $\epsilon > 2^{\mu - 1}$, we also have
		$T(2^{\mu}) = T(2^{\mu - 1}) + O(2^{\mu}\log2^{\mu})$. Hence, we have $T(2^{\mu}) = O(2^{\mu}\log2^{\mu})$. Finally, if we let the $\mu$ be the smallest integer such that $2^{\mu} \geq \epsilon$, the complexity of Algorithm \ref{alg:IFFTX_epsilon} is  $O(\epsilon\log\epsilon)$.

	\end{pf}
\end{lemma}

	Algorithm \ref{alg:IFFTX_epsilon} presents a general method for the $\epsilon$-points IFFT transform. However, it should be mentioned that, for a specific $\epsilon$, a further complexity reduction is possible. For example, Algorithm \ref{alg:ex_IFFT} is a special IFFT transform whose input size is not a power of 2.

\subsection{Encoding Algorithm for Arbitrary $(n,k)$ RS Codes}
	Now we turn to present the encoding algorithms for arbitrary $n,k$.
	Let $\epsilon = n - k$ and let $\mu$ be the smallest integer such that $2^{\mu}\geq \epsilon$. A RS code over finite field $\F_{2^m}$ of length $n$ and dimension $k$ is defined as
	\begin{equation*}
	\{(f(\omega_0), f(\omega_1), \dots, f(\omega_{n-1}))\mid \deg(f(x)) < 2^m - \epsilon, f(\omega_l) = 0, l = n, n + 1, \dots, 2^m - 1 \}.
	\end{equation*}
	Note that this is different from the classical definition in which an $(n,k)$ RS code is defined as 
	\begin{equation*}
	\{(f(\alpha_0), f(\alpha_1), \dots, f(\alpha_{n-1}))\mid \deg(f(x)) < k\}.
	\end{equation*}
	The reason is that we can derive fast encoding and decoding algorithm based on the new definition.
	
	Let $\omega_0, \dots, \omega_{\epsilon - 1}$ be check locations and $\omega_{\epsilon}, \dots, \omega_{n - 1}$ be message locations. We write $\mathbf{F} = (f(\omega_0), f(\omega_1), \dots, f(\omega_{2^m-1}))$.  According to Lemma \ref{lem:IFFT_property}, we then have
	\begin{equation}\label{app:eq1}
	\IFFT_{\bar{\X}}(\mathbf{F}_{0, \mu},\mu,0) + \IFFT_{\bar{\X}}(\mathbf{F}_{1, \mu},\mu,\omega_{2^{\mu}}) + \cdots + \IFFT_{\bar{\X}}(\mathbf{F}_{2^{m-\mu} - 1, \mu },\mu,\omega_{2^{m} - 2^{\mu}}) = \bar{\mathbf{f}}_{2^{m-\mu}-1, \mu}.
	\end{equation}
	Since $f(\omega_\epsilon),\ldots, f(\omega_{n-1})$ are the message symbols, so $\mathbf{F}_{1, \mu},\ldots,\mathbf{F}_{2^{m-\mu} - 1, \mu}$ in~\eqref{app:eq1} are known in advance. Note that if $\mathbf{F}_{l, \mu}$ is a zero vector, the corresponding IFFT transform in \eqref{app:eq1} is no need to perform. 
	Let
	\begin{equation*}
		\mathbf{w} = \IFFT_{\bar{\X}}(\mathbf{F}_{1, \mu},\mu,\omega_{2^{\mu}}) + \cdots + \IFFT_{\bar{\X}}(\mathbf{F}_{2^{m-\mu} - 1, \mu },\mu,\omega_{2^{m} - 2^{\mu}}).
	\end{equation*}
	Then we have
	\begin{equation*}
		\FFT_{\bar{\mathbb{X}}}(\bar{\mathbf{f}}_{2^{m-\mu}-1, \mu}, \mu, 0) = \mathbf{F}_{0, \mu} + \FFT_{\bar{\mathbb{X}}}(\mathbf{w}, \mu, 0).
	\end{equation*}
	Since $\deg(f(x)) < 2^m - \epsilon$, there are at least $\epsilon$ zeros in $\bar{\mathbf{f}}_{2^{m-\mu}-1, \mu}$. Hence, there are total $2^{\mu}$ unknowns in both side of the above equation and it can be solved uniquely. A method for solving this equation is presented in Algorithm \ref{alg:IFFTX_epsilon_spec} and its complexity is $O((2^{\mu} - \epsilon)\log(2^{\mu} - \epsilon))$. Since its proof is similar to that of Algorithm \ref{alg:IFFTX_epsilon}, so we omit it here. 
	
	A complete encoding algorithm is provided in Algorithm \ref{alg:enc_arbitrary}. Clearly, the encoding complexity is $O(n\log(n-k))$.
	
	\begin{breakablealgorithm}
		\caption{Special $\epsilon$-points IFFT Transform}
		\label{alg:IFFTX_epsilon_spec}
		\begin{algorithmic}[1]
			\REQUIRE
			$\{(F_{\epsilon}, F_{\epsilon + 1}, \dots, F_{2^{\mu} - 1}), 2^{\mu} - \epsilon, \mu, \beta\}$, where $\epsilon \leq 2^{\mu}$.
			\ENSURE
			$\{\bar{\mathbf{f}}, \mathbf{F}\}$
			such that $\bar{f}_{2^{\mu} - \epsilon}, \dots, \bar{f}_{2^{\mu} - 1} = 0$ and
			$\FFT_{\bar{\mathbb{X}}}(\bar{\mathbf{f}}, \mu, \beta) = \mathbf{F}$.
			
			\IF {$\mu = 0$} \RETURN $\{\bar{\mathbf{f}} = (F_0), \mathbf{F} = (F_0)\}$.
			\ENDIF
			\IF {$2^{\mu} - \epsilon \leq 2^{\mu - 1}$}
			\STATE Call Algorithm \ref{alg:IFFTX_epsilon_spec} with input 
			$\{(F_{\epsilon}, F_{\epsilon + 1}, \dots, F_{2^{\mu} - 1}), 2^{\mu} - \epsilon, \mu - 1, \omega_{2^{\mu - 1}} + \beta\}$
			to obtain $\{\bar{\mathbf{f}}_{0, \mu - 1}, \mathbf{F}_{1, \mu - 1}\}$.
			\STATE Call Algorithm \ref{FFTX} to obtain $\mathbf{F}_{0, \mu - 1} = \FFT_{\bar{\mathbb{X}}}(\bar{\mathbf{f}}_{0, \mu - 1},\mu - 1, \beta)$.
			\STATE Set $\bar{\mathbf{f}}_{1, \mu - 1} = \mathbf{0}$.
			\ELSE
			\STATE Call Algorithm \ref{IFFTX} to obtain $\mathbf{w} = \IFFT_{\bar{\mathbb{X}}}(\mathbf{F}_{1, \mu - 1}, \mu - 1, \omega_{2^{\mu - 1}} + \beta)$.
			\STATE Call Algorithm \ref{FFTX} to obtain $\mathbf{w}^{'} = \FFT_{\bar{\mathbb{X}}}(\mathbf{w}, \mu - 1, \beta)$.
			\STATE Let the input be
			$\{(F_{\epsilon}, \dots, F_{2^{\mu-1} - 1}) + (w_{\epsilon}^{'}, \dots, w_{2^{\mu-1} - 1}^{'}), 2^{\mu - 1} - \epsilon, \mu - 1, \beta \}$.
			Call Algorithm \ref{alg:IFFTX_epsilon} to get $\{\bar{\mathbf{f}}_{1, \mu - 1}, \mathbf{F}_{0, \mu - 1}^{'}\}$.
			\STATE Set $\bar{\mathbf{f}}_{0, \mu - 1} = \mathbf{w} + \frac{s_{\mu - 1}(\beta)}{s_{\mu - 1}(v_{\mu - 1})}\bar{\mathbf{f}}_{1, \mu - 1} + \bar{\mathbf{f}}_{1, \mu - 1}$ and $\mathbf{F}_{0,\mu - 1} = \mathbf{F}_{0, \mu - 1}^{'} + \mathbf{w}^{'}$.
			\ENDIF
			\RETURN $\{\bar{\mathbf{f}}, \mathbf{F}\}$.
		\end{algorithmic}
	\end{breakablealgorithm}

	\begin{breakablealgorithm}
		\caption{Encoding Algorithm for Arbitrary $(n,k)$ RS Codes}
		\label{alg:enc_arbitrary}
		\begin{algorithmic}[1]
			\REQUIRE
			The message symbols $f(\omega_{\epsilon}), \dots, f(\omega_{n-1})$ and $\epsilon$.
			\ENSURE
			The parity symbols $f(\omega_{0}), \dots, f(\omega_{\epsilon - 1})$.
			
			\STATE Find the smallest $\mu$ such that $2^{\mu} \geq \epsilon$.
			\STATE Compute the transforms
			$$\IFFT_{\bar{\mathbb{X}}}(\mathbf{F}_{l,\mu}, \mu, \omega_{l\cdot 2^{\mu}}), l = 1, 2, \dots, \lceil n/2^{\mu}\rceil - 1.$$
			\STATE Compute $\mathbf{w} = \sum_{l = 1}^{\lceil n/2^{\mu}\rceil - 1}\IFFT_{\bar{\mathbb{X}}}(\mathbf{F}_{l,\mu}, \mu, \omega_{l\cdot 2^{\mu}})$.
			\STATE Compute $\mathbf{w}^{'} = \FFT_{\bar{\mathbb{X}}}(\mathbf{w},\mu,0)$.
			\STATE Let the input be $\{f(\omega_\epsilon) + {w}_{\epsilon}^{'},\dots,f(\omega_{2^{\mu} - 1})+ {w}_{2^{\mu} - 1}^{'}, 2^{\mu}  - \epsilon, \mu, 0 \}$ and call Algorithm \ref{alg:IFFTX_epsilon_spec} to get $\{\bar{\mathbf{f}}_{2^{m-\mu} - 1}, \mathbf{F}_{0,\mu}^{'}\}$, where $\mathbf{F}_{0,\mu}^{'} = (f(\omega_{0}) + w_{0}^{'}, \dots, f(\omega_{2^{\mu} - 1}) + w_{2^{\mu} - 1}^{'})$.
			\STATE Compute $\mathbf{F}_{0,\mu} = \mathbf{F}_{0,\mu}^{'} + \mathbf{w}^{'}$, where $\mathbf{F}_{0,\mu} = (f(\omega_{0}), \dots, f(\omega_{2^{\mu} - 1})) $.
			\RETURN $f(\omega_{0}), \dots, f(\omega_{\epsilon - 1})$.
		\end{algorithmic}
	\end{breakablealgorithm}

\subsection{Decoding Algorithm for Arbitrary $(n,k)$ RS Codes}	
	We write the received vector as
	\begin{align*}
	\mathbf{r} = \mathbf{F} + \mathbf{e} &= (f(\omega_0), f(\omega_1), \dots, f(\omega_{2^m-1})) + (e_0, e_1, \dots, e_{2^m - 1})\\
	&=(r_0, r_1, \dots, r_{2^m - 1}),
	\end{align*}
	where $\mathbf{e}$ is the error pattern. Notice that only $f(\omega_0), \dots, f(\omega_{n - 1})$ are sent such that  $e_{n}, \dots, e_{2^m - 1} = 0$  and $f(\omega_{n}), \dots, f(\omega_{2^m-1})$ are equal to $0$. Thus, $r_{n}, \dots, r_{2^m - 1} = 0$. Let 
	$$E = \{\omega_l | e_l \neq 0\ \text{for}\ 0\leq l < n\}.$$
	The error locator polynomial is then defined by 
	$$\lambda(x) = \prod_{a\in E}(x - a).$$
	There exists a polynomial $r(x)\in\F_{2^m}[x]$ satisfying
	\begin{equation*}
		\deg(r(x)) < 2^m, r(\omega_l) = f(\omega_l) + e_l, l = 0, 1, \dots, 2^m - 1.
	\end{equation*}
	It follows that
	\begin{equation*}
		f(\omega_l)\lambda(\omega_l) = r(\omega_l)\lambda(\omega_l) \ \text{for}\ l = 0, 1, \dots, 2^m - 1.
	\end{equation*}
	Hence, the congruence
	\begin{equation*}
		f(x)\lambda(x) \equiv r(x)\lambda(x) \bmod{s_m(x)}
	\end{equation*}
	holds and we have
	\begin{equation}\label{eq:key_arb0}
		f(x)\lambda(x) = r(x)\lambda(x) + q(x)s_m(x)
	\end{equation}
	for some $q(x)$. According to our definition of an $(n,k)$ RS code, the degree of $f(x)$ is less than $2^m - \epsilon$, where $\epsilon = n - k$. Furthermore, $\deg(\lambda(x)) \leq \lfloor \epsilon / 2 \rfloor$. As $\deg(s_m(x)) = 2^m > \deg(r(x))$, one must have $\deg(q(x)) < \deg(\lambda(x))$.
	
	Let $\mu$ be the smallest integer such that $2^{\mu} \geq \epsilon$. Dividing $f(x)\lambda(x)$, $r(x)$, and $s_m(x)$ by $p_{2^m - 2^{\mu}}\bar{X}_{2^m - 2^{\mu}}(x)$ (which is a monic polynomial), we have
	\begin{align}
		f(x)\lambda(x) &= p_{2^m - 2^{\mu}}\bar{X}_{2^m - 2^{\mu}}(x)z^{'}(x) + \eta_f(x),\notag\\
		r(x) &= p_{2^m - 2^{\mu}}\bar{X}_{2^m - 2^{\mu}}(x)u(x) + \eta_{r}(x),\label{eq:u(x)}\\
		s_m(x) &= p_{2^m - 2^{\mu}}\bar{X}_{2^m - 2^{\mu}}(x)(s_\mu(x) + s_{\mu}(v_{\mu})) + \eta_s(x),\notag
	\end{align}
 	where the degrees of $\eta_f(x), \eta_r(x), \eta_s(x)$ are less than $\deg(p_{2^m - 2^{\mu}}\bar{X}_{2^m - 2^{\mu}}(x)) = 2^m - 2^{\mu}$.
 	Recall that 
 	\begin{equation*}
 		\deg(f(x)\lambda(x)) < 2^m - \epsilon + \deg(\lambda(x)) \mbox{ and }\deg(r(x))< 2^m.
 	\end{equation*}
 	This implies that
 	\begin{equation*}
 		\deg(z^{'}(x)) < 2^{\mu} - \epsilon + \deg(\lambda(x))\mbox{ and }\deg(u(x)) < 2^\mu.
 	\end{equation*}
	When we dividing both sides of \eqref{eq:key_arb0} by $p_{2^m - 2^{\mu}}\bar{X}_{2^m - 2^{\mu}}(x)$ and keeping the quotients, it becomes
	\begin{equation*}
		z^{'}(x) = u(x)\lambda(x) + q(x)(s_\mu(x) + s_{\mu}(v_{\mu})).
	\end{equation*}
	Let $z(x) = z^{'}(x) - s_{\mu}(v_{\mu})q(x)$. We obtain the equation
	\begin{equation}\label{eq:key0_n}
		z(x) = u(x)\lambda(x) + q(x)s_\mu(x),
	\end{equation}
	where $\deg(z(x)) < 2^{\mu}-\epsilon + \deg(\lambda(x))$ as $\deg(q(x)) < \deg(\lambda(x))$.
	
	Dividing $z(x)$, $u(x)$, and $s_{\mu}(x)$ by $\prod_{l = \epsilon}^{2^{\mu} - 1}(x - \omega_l)$, it follows that
	\begin{align*}
		z(x) &= \prod_{l = \epsilon}^{2^{\mu} - 1}(x - \omega_l)z_1(x) + \eta_z(x),\\
		u(x) &= \prod_{l = \epsilon}^{2^{\mu} - 1}(x - \omega_l)u_1(x) + \eta_{u}(x),\\
		s_{\mu}(x) &= \prod_{l = \epsilon}^{2^{\mu} - 1}(x - \omega_l) \prod_{j = 0}^{\epsilon - 1}(x - \omega_j),
	\end{align*}
	where $\deg(\eta_z(x)), \deg(\eta_{u}(x))$ are less than $\deg(\prod_{l = \epsilon}^{2^{\mu} - 1}(x - \omega_l)) = 2^{\mu} - \epsilon$. Evidently, one has
	\begin{align*}
		\deg(z_1(x)) &= \deg(z(x)) - (2^{\mu} - \epsilon) < \deg(\lambda(x)),\\
		\deg(u_1(x)) &= \deg(u(x)) - (2^{\mu} - \epsilon) < \epsilon.
	\end{align*}
	Dividing $\prod_{l = \epsilon}^{2^{\mu} - 1}(x - \omega_l)$ on both side of \eqref{eq:key0_n} and keeping the quotients, it becomes 
	\begin{equation}\label{eq:key_arb1}
		z_1(x) = u_1(x)\lambda(x) + q(x)\prod_{l = 0}^{\epsilon - 1}(x - \omega_l),
	\end{equation}
	where $\deg(z_1(x)) < \deg(\lambda(x))$. Note that if $\mathbf{r}$ is a codeword, then $\deg(r(x)) < 2^m - \epsilon$ and thus $\deg(u(x)) < 2^{\mu} - \epsilon$. This implies that $u_1(x) = 0$. Hence, $u_1(x)$ is the syndrome polynomial and \eqref{eq:key_arb1} is the key equation.
	
	Note that for solving the key equation, we only need $u_1(\omega_0), \dots, u_1(\omega_{\epsilon - 1})$. Now we present how to compute $u_1(\omega_0), \dots, u_1(\omega_{\epsilon - 1})$. 
	
	The polynomial $r(x)$ can be written as
	\begin{align*}
		r(x) = \sum_{l = 0}^{2^m - 1}\bar{r}_l\bar{X}_l(x) &= \sum_{l = 0}^{2^m - 2^\mu - 1}\bar{r}_l\bar{X}_l(x) + \sum_{j = 2^m - 2^{\mu}}^{2^m - 1}\bar{r}_j\bar{X}_j(x)\\
		&=\sum_{l = 0}^{2^m - 2^\mu - 1}\bar{r}_l\bar{X}_l(x) + p_{2^m-2^{\mu}}\bar{X}_{2^m - 2^{\mu}}(x)\sum_{j = 0}^{2^\mu - 1}\frac{\bar{r}_{j + 2^m - 2^{\mu}}}{p_{2^m-2^{\mu}}}\bar{X}_j(x),
	\end{align*}
	where the second identity holds since
	\begin{equation*}
		\bar{X}_{j + 2^m - 2^\mu}(x) = \bar{X}_{2^m - 2^\mu}(x)\bar{X}_{j}(x)\ \text{for}\ j < 2^{\mu}.
	\end{equation*}
	Because $\deg(\bar{X}_l(x)) < 2^m - 2^{\mu}$ for $l = 0,1, \dots, 2^m - 2^{\mu} - 1$, by \eqref{eq:u(x)},  we have 
	\begin{equation*}
		u(x) = \sum_{j = 0}^{2^\mu - 1}\frac{\bar{r}_{j + 2^m - 2^{\mu}}}{p_{2^m-2^{\mu}}}\bar{X}_j(x).
	\end{equation*}
	According to Lemma \ref{lem:IFFT_property}, the coefficient $(r_{2^m - 2^{\mu}}, \dots, r_{2^m - 1})$ can be computed by
	\begin{equation*}
		\sum_{l = 0}^{\lceil n / 2^{\mu}\rceil - 1}\IFFT_{\bar{\mathbb{X}}}(\mathbf{r}_{l,\mu}, \mu, \omega_{l\cdot 2^{\mu}}).
	\end{equation*}
	Therefore, the coefficients of $u(x)$ with respect to $\bar{\X}$ are obtained by
	\begin{equation}\label{eq:shorted_ux}
	\sum_{l = 0}^{\lceil n / 2^{\mu}\rceil - 1}\IFFT_{\bar{\mathbb{X}}}(\mathbf{r}_{l,\mu}, \mu, \omega_{l\cdot 2^{\mu}}) / p_{2^m-2^{\mu}}.
	\end{equation}
	Since
	\begin{equation}\label{eq:shorten_syndrome}
		u(x) = u_1(x) \prod_{l = \epsilon}^{2^{\mu} - 1}(x - \omega_l) + \eta_{u}(x),
	\end{equation}
	where $\deg(\eta_{u}(x)) < \deg(\prod_{l = \epsilon}^{2^{\mu} - 1}(x - \omega_l))$, it follows that $\eta_{u}(\omega_l) = u(\omega_l)$ for $l = \epsilon, \dots, 2^{\mu} - 1$. So $\eta_{u}(x)$ can be computed by Algorithm \ref{alg:IFFTX_epsilon_spec} once $u(\omega_l), l = \epsilon, \dots, 2^{\mu} - 1$ are obtained. Next, we evaluate $\eta_{u}$ at $\omega_0, \omega_1, \dots, \omega_{\epsilon - 1}$ by $\epsilon$-points FFT. Then $u_1(\omega_l), l = 0, 1, \dots, \epsilon - 1$ can be computed in a straightforward way according to \eqref{eq:shorten_syndrome}.  
	
	Once $u_1(\omega_l), l = 0, 1, \dots, \epsilon - 1$ are obtained. The key equation can be solved by the FDMA or FMA.
	Since FFT/IFFT transforms exist for any input size, it is easy to show the complexities of the FDMA and FMA are $O((n-k)^2)$ and $O((n-k)\log^2(n-k))$, respectively. 
	For example, let $\epsilon = 30$ and we solve the key equation by the FMA. The original problem is divided into two sub-problems of dimension $16$ and $14$. Solving the first sub-problem is straightforward since its size $16$ is equal to $2^4$. Next, we evaluate the solution to the first sub-problem at $\omega_0, \omega_1, \dots, \omega_{\epsilon}$ and update the second sub-problem. Then we can solve the updated second sub-problem inductively and evaluate its solution at $\omega_0, \omega_1, \dots, \omega_{\epsilon}$. Finally, we multiply these two solutions in frequency domain and do $(\epsilon + 1)$-input IFFT transform to get the desired result.
	
	After the error locator polynomial is obtained, finding the error locations can be done by
	\begin{equation}\label{eq:shorten_FFT_search}
		 \FFT_{{\bar{\mathbb{X}}}}(\bar{\lambda}, \mu, \omega_{l\cdot 2^{\mu}}), l = 0, 1, \dots, \lceil n/2^{\mu}\rceil - 1.
	\end{equation}
	
	According to the content in Subsection C, we have $f(\omega_l) - r(\omega_l) = q(\omega_l) / \lambda^{'}(\omega_l)$. If $\omega_l$ is a message location, then $q(\omega_l) = z_1(\omega_l)/\prod_{j = 0}^{\epsilon - 1}(\omega_l - \omega_j)$. Hence, Forney's formula for solving the error value is
	\begin{equation}\label{eq:shorten_Forney}
		f(\omega_l) - r(\omega_l) = \frac{z_1(\omega_l)}{\prod_{j = 0}^{\epsilon - 1}(\omega_l - \omega_j)\lambda^{'}(\omega_l)}.
	\end{equation}
	Note that $\prod_{j = 0}^{\epsilon - 1}(\omega_l - \omega_j)$ can be computed in advance and stored for $l = \epsilon, \dots, n - 1$.
	
	\begin{algorithm}[h]
		
			\caption{{Decoding Algorithm for Arbitrary $(n,k)$ RS Codes}}
			\label{alg:decoding_arbitrary}
			\begin{algorithmic}[1]
				\REQUIRE
				Received vector $\mathbf{r} = \mathbf{F} + \mathbf{e}$.
				\ENSURE
				The codeword $\mathbf{F}$.
				\STATE Compute the syndrome polynomial $u(x)$ according to \eqref{eq:shorted_ux}.
				\STATE Evaluate $u(x)$ at points $\omega_0, \omega_1, \dots, \omega_{2^{\mu}-1}$ by Algorithm \ref{FFTX}.
				\STATE Given $\eta_{u}(\omega_{l}) = u(\omega_{l})$ for $l = \epsilon, \dots, 2^{\mu} - 1$, call Algorithm \ref{alg:IFFTX_epsilon_spec} to get $\eta_{u}(x)$.
				\STATE Evaluate $\eta_{u}(x)$ at $\omega_0, \dots, \omega_{\epsilon}$ by $\epsilon$-points FFT and compute $u_1(\omega_l)$ for $l = 0, 1, \dots, \epsilon - 1$ according to \eqref{eq:shorten_syndrome};
				\STATE Given $\phi_i(W(x), N(x)) = u_1(\omega_{i-1})W(\omega_{i-1}) + N(\omega_{i-1}), i = 1, 2, \dots, \epsilon$, compute the error locator polynomial $\lambda(x)$ and the error evaluator polynomial $z_1(x)$ by Algorithms \ref{alg:FDMA} or \ref{alg:FMA}.
				\STATE Find the error locations by \eqref{eq:shorten_FFT_search}.
				\STATE Compute the error pattern $\mathbf{e}$ by \eqref{eq:shorten_Forney}.
				\RETURN
				$\mathbf{r} + \mathbf{e}$.
			\end{algorithmic}
		
	\end{algorithm}

	A complete description of decoding algorithm for arbitrary $(n,k)$ RS codes is shown in Algorithm \ref{alg:decoding_arbitrary}. Obviously, the decoding complexity for an arbitrary $(n, k)$ RS code is $O(n\log(n-k) + (n-k)\log^2(n-k))$.

\end{appendices}


\ifCLASSOPTIONcaptionsoff
  \newpage
\fi
\bibliographystyle{IEEEtran}
\bibliography{IEEEabrv,refs_EDITED_QCd}

\end{document}